\documentclass[a4paper,11pt]{article}
\usepackage{jheppub} 
\usepackage{lineno}
\usepackage{nicefrac}

\newcommand{\onech}[2]{\big[{\textstyle{{#1 \atop #2}}}\big]}
\newcommand{\twoch}[4]{\Big[{\textstyle{{#1 \atop #3} {#2 \atop #4}}}\Big]}

\title{\boldmath An obstruction to generalised discrete torsion in string theory}

\author[a]{Carlo Angelantonj,}
\author[b,c]{Giorgio Leone}
\author[d]{and Diego Perugini}
\affiliation[a]{Dipartimento di Fisica , University of Torino, \\ INFN Sezione di Torino,\\
Via P. Giuria 1, 10125 Torino -- Italy}
\affiliation[b]{Dipartimento di Fisica e Astronomia ``Galileo Galilei'', University of Padova, \\ Via F. Marzolo 8, 35131 Padova -- Italy}
\affiliation[c]{NFN Sezione di Padova, \\ Via F. Marzolo 8, 35131 Padova -- Italy}
\affiliation[d]{Institut de Physique Th\'eorique, Universit\'e Paris Saclay, CEA, CNRS, \\ Orme des Merisiers,
91191 Gif-sur-Yvette CEDEX -- France}

\emailAdd{carlo.angelantonj@unito.it, giorgio.leone@unipd.it, diego.perugini@ipht.fr}

\abstract{We reconsider the possibility of turning on generalised discrete torsion in orbifold compactifications, whereby different discrete torsion phases are attached to different fixed points of a given twisted sector. By an explicit inspection of the genus one and genus two vacuum amplitudes for the $T^6/\mathbb{Z}_2^2$ and $T^7/\mathbb{Z}_2^3$ cases, we show that this is not possible in String Theory. In fact, this generalised discrete torsion, although modular invariant at genus one, leads to an inconsistent interpretation of the spectrum which, in turn, implies a loss of modular invariance on genus-two Riemann surfaces. Our finding is in agreement with a recent classification of allowed topological deformations in equivariant cohomology, which does not include this possibility.}

\begin{document}
\maketitle
\flushbottom

\newpage

\section{Introduction}
\label{sec:intro}

Orbifolds \cite{Dixon:1985jw,Dixon:1986jc} are a special playground for studying non-trivial compactifications of string theory. In particular, toroidal orbifolds are flat almost everywhere up to, possibly, a finite number of loci where the curvature is concentrated. As a result, they admit an exact conformal field theory (CFT) description, and strings can consistently propagate on such singular spaces since they provide the \emph{tools} to desingularise them.

Albeit the standard lore is that modular invariance is so powerful to fully determine the spectrum of closed oriented strings, this is not true in general orbifold compactifications. Indeed, it was shown by Vafa \cite{Vafa:1986wx} that, in some cases, one can introduce suitable phases in the one-loop partition function, which are fully compatible with modular invariance. Clearly, this freedom affects the light and massive spectrum and, in turn, modifies the topology of the resulting compactification manifold. The prototype example is the $T^6/\mathbb{Z}_2 \times \mathbb{Z}_2$ orbifold of \cite{Vafa:1994rv}. Although a faithful representation of the orbifold group on the string coordinates completely fixes the projection of untwisted strings, \emph{i.e.} those which are periodic on the world-sheet, this is not the case for the twisted ones. For instance, if we denote by $g$ and $h$ two elements of $\mathbb{Z}_2 \times \mathbb{Z}_2$, while the $g$ projection of the $g$ twisted sector is uniquely dictated by modular invariance, its $h$ projection is ambiguous. This is related to the fact that this orbifold admits a fully independent modular orbit which completes the projection in the twisted sectors, and supports a sign ambiguity called \emph{discrete torsion}. The resulting spaces are Calabi-Yau manifolds with Hodge numbers $(h^{1,1},h^{1,2}) = (51,3)$ for a positive discrete torsion, and $(h^{1,1},h^{1,2}) = (3,51)$ for negative discrete torsion \cite{Vafa:1994rv}. These choices correspond to different desingularisations of the 48 fixed loci: on the one hand, when the discrete torsion is positive, the singularities are resolved adding 48 smooth 2-cycles, each contributing one unit to $h^{1,1}$. On the other hand, when the discrete torsion is negative, the singularities are deformed adding 48 smooth 3-cycles, each contributing one unit to $h^{1,2}$.

In classical geometry, it is actually possible to simultaneously resolve and deform different singularities of a given orbifold. In this way, starting from $T^6 /\mathbb{Z}_2^2$ and $T^7 / \mathbb{Z}_2^3$, Joyce was able to construct families of smooth Calabi-Yau and special $G_2$ holonomy spaces with varying cohomologies \cite{Joyce:1998en, JoyceBook}. These constructions led Gaberdiel and Kaste \cite{Gaberdiel:2004vx} to propose a generalisation of the concept of discrete torsion. Their main idea was to treat independently the singular loci, by assigning different phases to each of them, thus implementing in string theory a proposal of Joyce to perform different desingularisations of the fixed loci. Although, it was shown in \cite{Gaberdiel:2004vx} that this generalised (point-wise) discrete torsion passes the test of one-loop modular invariance, it actually fails at building a consistent string background. In fact, a \emph{bona fide} CFT not only has to respect one-loop modular invariance, but it should also provide a sounding spectrum of excitations. Moreover, in string theory, we are instructed to sum over Riemann surfaces of arbitrary genus, and thus higher-genus modular invariance is as important as the one-loop counterpart. As we shall see, the proposal of \cite{Gaberdiel:2004vx} fails to comply with both requirements.

A twisted vacuum is built out of twist fields which carry a non-trivial U(1) charge \cite{Dixon:1986qv}. In this case, in the left-moving sector, for any twisted direction, there are twisted fields $\sigma_a $ and $\sigma_{\dot a}$ with positive and negative charge, respectively, and similarly for the right-moving sector. A twisted vacuum can then be neutral, $\sigma_a \tilde\sigma_{\dot a}|0,\tilde 0\rangle$ or $\sigma_{\dot a} \tilde\sigma_a |0,\tilde 0\rangle$, or carry a non-trivial charge, $\sigma_a \tilde\sigma_c |0,\tilde 0\rangle$ or $\sigma_{\dot a} \tilde\sigma_{\dot c} |0,\tilde 0\rangle$. Which of the twisted ground states is selected by the compactification depends entirely on how the orbifold group $\varGamma$ acts on the untwisted sector. Once this choice is made, the discrete torsion of \cite{Vafa:1986wx} simply determines the parity of these ground states under $\varGamma$, and thus the contribution of the twisted sectors to the cohomology of the compactification manifold. Actually, an orbifold admitting discrete torsion has several twisted sectors, and the modular transformations of the elements of the independent orbit correlate the parity of the twisted ground states in all sectors. Any attempt of generalisation has to implement a well defined projection of such ground states, thus providing a consistent operator interpretation of the associated CFT. In a given twisted sector, the generalised discrete torsion of \cite{Gaberdiel:2004vx} assigns different parities to different fixed loci, so that each contributes differently to the cohomology. Although this option is, at first sight, innocuous, upon modular transformations, it yields an inconsistent projection in the other twisted sectors, since the independent modular orbit would now project ground states and excitations which do not belong to the Hilbert space. This inconsistency is hard to see at a generic point of moduli space. In fact, at these points, the partition function simply counts the number of degrees of freedom, and is not able to distinguish between the ground states $\sigma_a \tilde\sigma_{\dot a}|0,\tilde 0\rangle$ and $\sigma_a \tilde\sigma_c |0,\tilde 0\rangle$. To check whether the independent modular orbit is indeed projecting the correct states, it would be necessary to turn on chemical potentials (or background magnetic fields) which are sensible to the U(1) charge. Although, this may be complicated to do at a generic point of moduli space, Rational CFT's (RCFT's) provide a powerful way to keep track of the various quantum numbers, since all excitations are packed into characters whose label is directly linked to their U(1) charge. The generalised discrete torsion of \cite{Gaberdiel:2004vx} can be straightforwardly implemented also in this case and, although compatible with modular invariance, it does not provide a consistent particle interpretation. To be specific, while it can be chosen to build a well defined spectrum in, say, the $g$ twisted sector, the unambiguous modular transformations of the characters, brings in inconsistencies in the different $h$ twisted sector, where the ground state is selected to be
\begin{equation}
{\mathcal Z} \onech{h}{1} \supset \sigma^3_a \tilde \sigma^3_a \, \sigma^4_{\dot a } \tilde \sigma^4_{\dot a} \, \sigma^5_{\dot a } \tilde \sigma^5_{\dot a} \, \sigma^6_a \tilde \sigma^6_a \, |0,\tilde 0\rangle\,,
\end{equation}
while the independent modular orbit would project the state
\begin{equation}
\sigma^3_a \tilde \sigma^3_a \, \sigma^4_{\dot a } \tilde \sigma^4_{ a} \, \sigma^5_{\dot a } \tilde \sigma^5_{\dot a} \, \sigma^6_a \tilde \sigma^6_a \, |0,\tilde 0\rangle\,,
\end{equation}
which differs in the nature of the $\tilde\sigma^4_a$ twist field.

In no way this rational analysis is of limited validity. In fact, the spectrum provides marginal deformations which can continuously connect the RCFT to any irrational CFT within the same moduli space, thus preserving our conclusions.

For RCFT's it is known that one-loop modular invariance and a consistent operator interpretation automatically guarantee  the invariance under reparametrisations of higher genus surfaces  \cite{Antoniadis:1986rn, Moore:1988qv,  Gaberdiel:2010jf}. It is natural to believe that this connection holds also for generic CFT, so one could expect that the generalised discrete torsion of \cite{Gaberdiel:2004vx}, although compatible with one loop modular invariance, should fail to respect it at higher orders. To support this argument, we explicitly compute the genus-two vacuum energy, at a generic point in moduli space, and find that only the discrete torsion of \cite{Vafa:1986wx} is compatible with two-loop modular invariance and factorisation.

Recently, Boyle Smith and Tachikawa \cite{BoyleSmith:2026oay} have also discussed the consistency of the generalised discrete torsion of \cite{Gaberdiel:2004vx}. Their analysis is based on the study of the  topological deformations of the $T^6/\mathbb{Z}_2^2$ and $T^7/\mathbb{Z}_2^3$ orbifolds classified by the equivariant cohomology group $H^2_\varGamma (T^n , \text{U} (1))$. Although, these deformations, consistent by construction, allow to build CY and $G_2$ manifolds with varying cohomology, they \emph{do not} give a consistent implementation of the proposal of \cite{Gaberdiel:2004vx}. In short, the allowed deformations of \cite{BoyleSmith:2026oay} are \emph{conventional} discrete torsions between shifts and/or twists associated to global $B$-field fluxes \cite{Sharpe:2000ki, Sharpe:2003cs}. In no way they provide a point-wise generalisation of the discrete torsion of Vafa between pure twists \cite{Vafa:1986wx}.
In this paper we shall focus of the proposal of \cite{Gaberdiel:2004vx}, while in the forthcoming publication \cite{CDGtoappear} we shall study and interpret the topological deformations classified by the equivariant cohomology group for generic $T^6/\mathbb{Z}_N \times \mathbb{Z}_M$ orbifolds.

Our results and those of \cite{BoyleSmith:2026oay} leave unanswered the question about a stringy realisation of the freedom one has in resolving or deforming the orbifold singularities in classical algebraic geometry \cite{Joyce:1998en, JoyceBook}.

The paper is organised as follows. In Section \ref{sec:attgdt} we review the proposal of \cite{Gaberdiel:2004vx} to define a generalised discrete torsion. In Section \ref{sec:Oneloop} we build the one loop partition functions for the $T^6/\mathbb{Z}_2^2$ and $T^7/\mathbb{Z}_2^3$ orbifolds at the $\text{SU}(2)^6$ and $\text{SU}(2)^7$ points, respectively. We introduce the generalised discrete torsion and show that it provides an inconsistent operator interpretation. In Section \ref{sec:genus2} we explicitly compute the genus-two vacuum energy for the $T^6/\mathbb{Z}_2^2$ orbifold, study its factorisation limit and prove that it can only support the standard discrete torsion of Vafa. We also provide strong evidence that the same results hold for the $T^7/\mathbb{Z}_2^3$ case.
Finally, three Appendices collect useful background material on genus-$g$ theta functions and details on one and two-loop vacuum amplitudes.

\section{An attempt at generalised discrete torsion}\label{sec:attgdt}

An instructive and simple example to understand the role of discrete torsion \cite{Vafa:1986wx} and its possible generalisation \cite{Gaberdiel:2004vx} is the string compactification on the $T^6/\mathbb{Z}_2 \times \mathbb{Z}_2$ orbifold. More general $\mathbb{Z}_N \times \mathbb{Z}_M$ cases, with $N$ and $M$ not coprime, add minor complications which, however, do not bring any conceptual novelty to the discussion. The orbifold generators $g$ and $h$ act on the real $T^6$ coordinates $x^i$ ($i=1,\ldots ,6$) as 
\begin{equation}
    g: \ (-,-,-,-,+,+)\,, \qquad h:\ (+,+,-,-,-,-)\,,
    \label{Z2Z2generators}
\end{equation}
where each entry defines the associated eigenvalue along the $i$-th direction. The other element of the orbifold group is $f=gh$.

\begin{figure}
\centering
\includegraphics[scale=0.5]{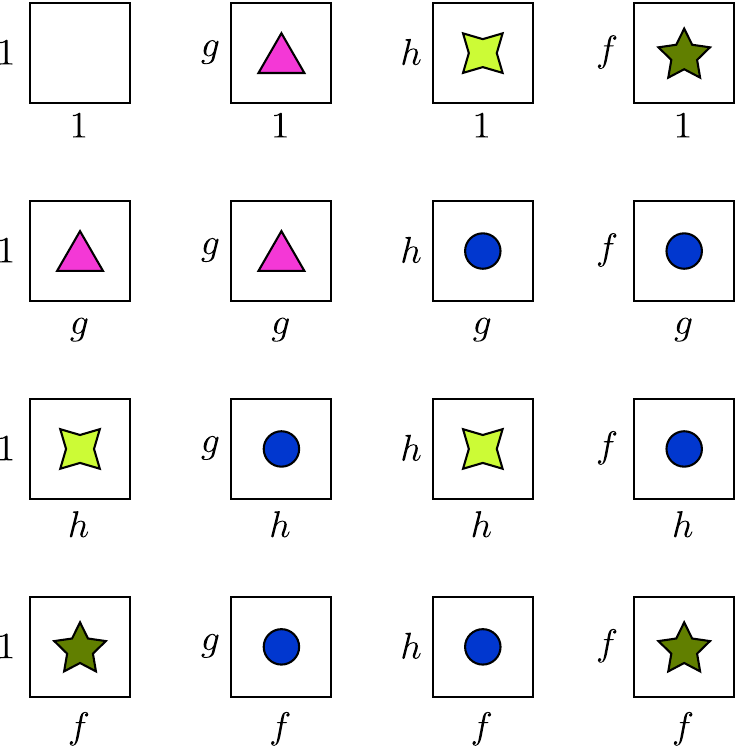}
\caption{The structure of the $T^6/\mathbb{Z}_2\times \mathbb{Z}_2$ orbifold. The boxes represent the world-sheet torus, where time flows vertically and the closed string extends horizontally. Boxes decorated with the same symbol are related by modular transformations and, thus, belong to the same modular orbit. The boxes with a blue disc can support discrete torsion.}
\label{Z2Z2}
\end{figure}

The spectrum of string excitations on this space is, as usual, encoded in the one-loop partition function
\begin{equation}
    {\mathcal Z} = \frac{1}{4} \sum_{\alpha,\beta} \, \text{tr}_{\mathcal{H}_{\alpha}} \left( \beta \, q^{L_0-c/24} \, \bar q^{\bar L_0-c/24} \right)\equiv \sum_{\alpha,\beta} {\mathcal Z}\onech{\alpha}{\beta}\,,
    \label{pfZ2Z2generic}
\end{equation}
with $\alpha , \beta \in \{ \mathbf{1}, g , h ,f \}$  generic orbifold elements, and $\mathcal{H}_{\alpha}$ the Hilbert space associated to the closed strings twisted by $\alpha$, \textit{i.e.} closed strings which are periodic modulo the action of the group element, ${\boldsymbol X} (\sigma +\pi , \tau ) = \alpha \cdot {\boldsymbol X} (\sigma , \tau)$. Although ${\mathcal Z}$ is, as required, invariant under the action of the modular group $\varGamma = \text{PSL}(2;\mathbb{Z})$, each individual term in \eqref{pfZ2Z2generic} is not, since
\begin{equation}
    S:\ {\mathcal Z}\onech{\alpha}{\beta} \to {\mathcal Z}\onech{\beta}{\alpha^{-1}} \qquad \text{and}\qquad
    T:\ {\mathcal Z}\onech{\alpha}{\beta} \to {\mathcal Z}\onech{\alpha}{\alpha\beta}\,,
    \label{modularconnection}
\end{equation}
with $S$ and $T$ the usual generators of $\varGamma$. It is easy to realise that, in this $\mathbb{Z}_2 \times \mathbb{Z}_2$ example, all amplitudes ${\mathcal Z}\onech{\alpha}{\beta}$ form five independent modular orbits each generated by 
\begin{equation}
    {\mathcal Z}\big[{\textstyle{1 \atop \alpha}}\big] \ \text{with}\  \alpha \in \{\mathbf{1},g,h,f\}\,, \qquad \text{and, for instance,} \qquad {\mathcal Z}\big[{\textstyle{g \atop h}}\big]\,,
\end{equation}
as depicted in figure \ref{Z2Z2}.
Therefore, one-loop modular invariance does not uniquely determine ${\mathcal Z}$, since each independent orbit may appear with an arbitrary sign. However, a faithful action of the orbifold generators on the untwisted string coordinates uniquely fixes the ${\mathcal Z}\big[{\textstyle{1 \atop \alpha}}\big]$'s while leaving a complete freedom for a sign $\epsilon = \pm 1$ associated to the orbit generated by ${\mathcal Z}\big[{\textstyle{g \atop h}}\big]$. This is precisely the discrete torsion introduced by Vafa \cite{Vafa:1986wx}.

Another characteristic of orbifolds is that twisted Hilbert spaces are typically degenerate, and come in multiple independent copies, each associated to a different fixed locus
\begin{equation}
    {\mathcal H}_\alpha = \bigoplus_{r} {\mathcal H}_\alpha^r\,.
\end{equation}
Indeed, in our case, each $\alpha$-twisted sector has a 16-fold multiplicity associated to the 16 $T^2$'s fixed under the action of $\alpha$ and, schematically,
\begin{equation}
\begin{aligned}
   \frac{1}{4} \sum_{\beta} \text{tr}_{\mathcal{H}_{\alpha}}  \beta q^{L_0-c/24} \, \bar q^{\bar L_0-c/24} &= \sum_{r=1}^{16} \frac{1}{4} \sum_{\beta} \text{tr}_{\mathcal{H}_{\alpha}^r}  \beta q^{L_0-c/24} \, \bar q^{\bar L_0-c/24} 
\\
&\equiv 16 \left[ \frac{1}{4}\sum_{\beta= 1,\alpha} \hat {\mathcal Z} \big[{\textstyle{\alpha \atop \beta}}\big] + \epsilon\, \frac{1}{4} \sum_{\beta\neq 1,\alpha } \hat {\mathcal Z} \big[{\textstyle{\alpha \atop \beta}}\big]  \right]\,,
\end{aligned}
\end{equation}
where, in the second line, the terms within the brackets encode the (identical) spectrum associated to each fixed locus, while the overall factor takes into account their multiplicity. 
Notice that the conventional discrete torsion \cite{Vafa:1986wx}, $\epsilon = \pm 1$,  induces the same projection on all fixed points, thus preserving the degeneracy.

A natural question is whether one can actually associate a different $\epsilon$ at each fixed locus, thus inducing a different projection on the states in ${\mathcal H}_\alpha^r$ for different $r$. After all, Joyce suggested that, in classical geometry, one can proceed with different desingularisations (resolution or deformation) for each fixed $T^2$ in $\mathbb{Z}_2^n$ orbifolds, thus building inequivalent families of Calabi-Yau and $G_2$ manifolds \cite{Joyce:1998en, JoyceBook}.

At first sight, this \emph{local} version of discrete torsion seems to be possible in string theory since the independent orbit generated by $\hat {\mathcal Z} \onech{g}{h} = {\mathcal Z} \onech{g}{h} /16$, and encoding the projected spectrum of a given fixed locus, is still modular invariant. This led Gaberdiel and Kaste \cite{Gaberdiel:2004vx} to introduce a generalised, point-wise, version of discrete torsion, where each fixed locus is equipped with an independent $\epsilon_r =\pm 1$. The new partition partition function for the $T^6/\mathbb{Z}_2^2$ orbifold thus reads
\begin{equation}
    \frac{1}{4} \sum_{\beta} \text{tr}_{\mathcal{H}_{\alpha}}  \left( \beta\,  q^{L_0-c/24} \, \bar q^{\bar L_0-c/24} \right) = \sum_{r=1}^{16} \left[ \frac{1}{4}\sum_{\beta= 1,\alpha} \hat {\mathcal Z} \onech{\alpha}{\beta} + \epsilon_r\, \frac{1}{4} \sum_{\beta\neq 1,\alpha} \hat {\mathcal Z} \onech{\alpha}{\beta}  \right] \, . \label{eq:gdt}
\end{equation}
The sixteen-fold degeneracy of the $\alpha$ twisted sector is now (partially) lifted since each fixed locus may now support a different spectrum, depending on the different choices of $\epsilon_r$. For instance, if $n_\pm$ counts the number of positive/negative discrete torsions $\epsilon_r$ with, of course, $n_+ + n_- =16$, the partition function in a generic $\alpha$-twisted sector reads
\begin{equation}
 (n_+ + n_-) \frac{1}{4} \left[ \hat {\mathcal Z} \onech{\alpha}{1} +  \hat {\mathcal Z} \onech{\alpha}{\alpha} \right]+ (n_+ - n_-) \frac{1}{4}\left[ \hat {\mathcal Z} \onech{\alpha}{\beta} +  \hat {\mathcal Z} \onech{\alpha}{\alpha\beta} \right]\,.
 \label{GDTexpl}
 \end{equation}
Following \cite{Joyce:1998en, JoyceBook, Gaberdiel:2004vx}, fixed loci with positive or negative discrete torsion $\epsilon_r$ are either resolved or deformed and thus contribute one unit to the Hodge number $h^{1,1}$ or $h^{1,2}$, respectively. Together with the contribution $(h^{1,1},h^{1,2})=(3,3)$ from the untwisted sector, this would result into a family of $\text{CY}_3$ manifolds with $(h^{1,1},h^{1,2})= (3 + 3 n_+ , 3 + 3 n_-) = (51- 3 n_- , 3 + 3 n_-)$, $n_- = 0 ,\ldots , 16$.

\begin{figure}
\centering
\includegraphics[scale=0.35]{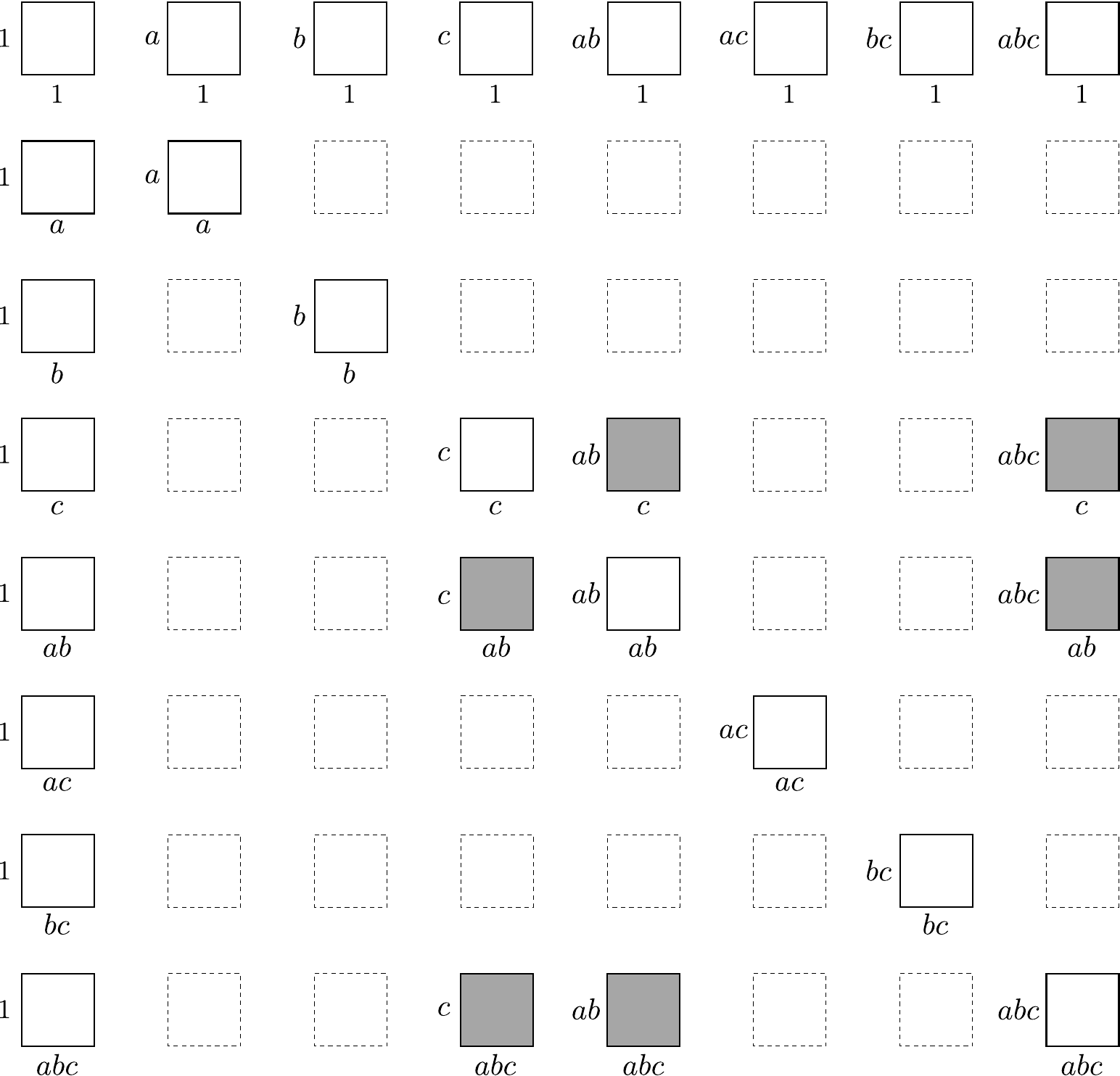}
\caption{The structure of the $T^7/\mathbb{Z}_2^3$ orbifold describing the singular limit of the $G_2$ holonomy 7-manifold. The boxes represent the world-sheet torus, where time flows vertically and the closed string extends horizontally. Dashed boxes denote orbifold amplitudes which vanish identically due to the combined action of shifts and rotations. The grey boxes denote the only non-trivial modular orbit which may support discrete torsion.}
\label{G2}
\end{figure}

A similar analysis can be repeated for the seven-dimensional  manifold of special $G_2$ holonomy, built as the $T^7 /\mathbb{Z}_2^3$ orbifold \cite{Joyce:1998en, JoyceBook}, which was originally discussed in  \cite{Acharya:2001gy, Font:2010sj} as a consistent string background.  The three generators are
\begin{equation}
\begin{aligned}
a &= (-,-,-,-,+,+,+ | 0,0,0,0,0,0,0)\,,
\\
b &= (-,-,+,+,-,-,+ | 0,\nicefrac{1}{2},0,0,0,0,0)\,,
\\
c &= (-,+,-,+,-,+,- | \nicefrac{1}{2},0,0,0,0,0,0)\,,
\end{aligned}\label{G2gen}
\end{equation}
where the signs denote the reflection along the corresponding real coordinate of $T^7$, while the $\nicefrac{1}{2}$ indicates an order-two shift along the associated coordinate, $x\to x+\pi$ with $x\in (0,2\pi )$. Given the presence of twists and shifts, many contributions to the partition function vanish, and one is left with the non-trivial amplitudes ${\mathcal Z} \onech{\alpha}{\beta}$ schematically represented in figure \ref{G2}. For instance, the action of $b$ in the $a$ twisted sector exchanges the fixed points with $x^2 =0$ and $x^2 = \pi$, thus yielding a vanishing trace over ${\mathcal H}_a$.

Moreover, the sectors twisted by $ab$,  $ac$, $bc$ and $abc$ only contain massive excitations, since the presence of the shift in the generators implies half-integer winding modes in these sectors. Therefore, the discrete torsion associated to the modular orbit generated by ${\mathcal Z} \onech{c}{ab}$ can only affect the desingularisation of the $c$-twisted sector. Following \cite{Gaberdiel:2004vx} and allowing for independent discrete torsions for each of the eight $c$-fixed points, as in eq. \eqref{GDTexpl}, one finds the Betti numbers
\begin{equation}
(b_0 \, \ldots , b_7) = (1,0,8+\ell , 47 - \ell , 47 - \ell , 8+\ell , 0 , 1) \qquad \text{for}\qquad \ell =0,1, \ldots , 8\,,
\end{equation} 
describing the family of $G_2$ manifolds of \cite{Joyce:1998en, JoyceBook}.

Although, this generalised discrete torsion \cite{Gaberdiel:2004vx} preserves one-loop modular invariance and reproduces the results of Joyce in classical geometry \cite{Joyce:1998en, JoyceBook}, an attentive study reveals that it does not provide a valid operator interpretation in the CFT and, moreover, fails two-loop modular invariance. This is what we will show in the following Sections. Our results are in agreement with a recent analysis \cite{BoyleSmith:2026oay} where it is shown that the allowed (discrete) topological deformations of an orbifold $M/\varGamma$ are classified by the equivariant cohomology classes in $H^2_\varGamma (M; \text{U} (1))$, which do not include the generalised (point-wise) discrete torsion of \cite{Gaberdiel:2004vx}.\footnote{Notice that the discrete deformations of \cite{BoyleSmith:2026oay} definitely do not include the point-wise generalised discrete torsion of \cite{Gaberdiel:2004vx}, although in the introduction the authors of \cite{BoyleSmith:2026oay} comment on this possible link.}

\section{Genus-one modular invariance and the consistency of CFT's} \label{sec:Oneloop}

The simple world-sheet analysis of \cite{Gaberdiel:2004vx} is actually missing crucial details of the twisted Hilbert spaces. In fact, the partition function \eqref{pfZ2Z2generic} only counts the total number of degrees of freedom at each mass level, regardless of the charges carried by the various states. For instance, in the $T^6 / \mathbb{Z}_2^2$ orbifold the contribution of the world-sheet bosons to the various amplitudes can be expressed in terms of Jacobi theta constants as
\begin{equation}
\begin{aligned}
\hat {\mathcal Z} \big[{\textstyle{\alpha \atop 1}}\big] &\sim \left| \frac{\eta^2}{\vartheta_4^2}\right|^2 \, \varLambda_{(2,2)}^{(k)} \simeq 
1 +  4\, q^{1/2} + 4\, {\bar q}^{1/2} +  16\, (q\, \bar q)^{1/2} + 12\, q + 12\, \bar q  +\ldots
\,,
\\
\hat {\mathcal Z} \big[{\textstyle{\alpha \atop \alpha}}\big] &\sim \left| \frac{\eta^2}{\vartheta_3^2}\right|^2 \, \varLambda_{(2,2)}^{(k)}\simeq
1 -  4\, q^{1/2} - 4\, {\bar q}^{1/2} +  16\, (q\, \bar q)^{1/2} + 12\, q + 12\, \bar q  +\ldots
\,,
\\
\hat {\mathcal Z} \big[{\textstyle{\alpha \atop \beta}}\big] &\sim \left| \frac{ 2 \eta^3}{\vartheta_2\vartheta_3 \vartheta_4 }\right|^2 = 1\,,
\end{aligned}\label{thetapf}
\end{equation}
for any $\alpha \neq \mathbf{1}$ and $\beta \neq \mathbf{1}, \alpha$. From this expression one reads the existence of a single (twisted) vacuum associated to each fixed point, but it is impossible to extract information about the nature of its twist fields. For this $\mathbb{Z}_2^2$ orbifold, the twist fields are built out of suitable products of $\sigma_a \tilde \sigma_b$, $\sigma_{\dot a} \tilde \sigma_{b}$ and their conjugates. The twist fields $\sigma_a$ and $\sigma_{\dot a}$ carry opposite $\text{U} (1)$ charges which, however, are not spelled out in eq. \eqref{thetapf}. Resolving this ambiguity in irrational CFT's is referred to as \emph{fixed-point resolution problem} \cite{Fuchs:1995tq, Fuchs:1996dd, Fuchs:1996rq}, and is usually challenging.

To overcome this ambiguity, one might introduce suitable chemical potentials or, alternatively, sit at special points in moduli space, where the CFT becomes rational, the partition function is written in terms of its characters, and the resolution of the fixed-point ambiguity is built in. This is precisely what we shall resort to. Following \cite{Gaberdiel:2004vx, BoyleSmith:2026oay}, we shall only consider the contribution of the world-sheet bosons. They contain all information about fixed points and twist fields, and suffice to draw our conclusions. Including world-sheet fermions and/or gauge degrees of freedom, in the case of the heterotic strings, is straightforward but does not bring any extra insight in our analysis.

Without loss of generality, we assume that the compactification torus $T^d$ ($d=6$ for $\text{CY}_3$ and $d=7$ for the $G_2$ manifolds) factorises into the product of $d$ circles, each with self-dual radius $R=\sqrt{\alpha'}$. The Narain lattice associated to each compact direction can then be written in terms of level-one SU(2) characters,
\begin{equation}
\varLambda_{(d,d)} = \left(\varLambda_{(1,1)} (R=\sqrt{\alpha '} )\right)^d\,, \qquad
\varLambda_{(1,1)} (R=\sqrt{\alpha '})= |\chi_0 |^2 + |\chi_{\nicefrac{1}{2}}|^2 \,, \label{sixtorus}
\end{equation}
where $\chi_0 (q)$ ($\chi_{\nicefrac{1}{2}} (q)$) is associated to the conjugacy class of (half-)integer-spin representations. A reflection along this compact coordinate,  $x\to - x$, breaks $\text{SU} (2) \to \text{U} (1)_4$, where $\text{U} (1)_4$ is realised in terms of a free compact boson at radius $R=\sqrt{4 \alpha '}$, with characters $\xi_a (q)$, $a=-3,\dots , 4$. One can decompose
\begin{equation}
\chi_0 = \xi_0 + \xi_4\,, \qquad \chi_{\nicefrac{1}{2}} = \xi_2 + \xi_{-2}\,, \label{su2u1}
\end{equation}
while the reflection acts as
\begin{equation}
|\xi_0 + \xi_4|^2 \to |\xi_0 - \xi_4|^2 \,, \qquad \text{and} \qquad |\xi_2 + \xi_{-2}|^2 \to |\xi_2 - \xi_{-2}|^2 \,. \label{parityaction}
\end{equation}
Details on these characters and their modular properties are given in Appendix \ref{AppA}, but it is worth to stress here that the index $a$ labels their $\text{U} (1)$ charge. Therefore, writing the partition function in terms of the $\xi_a$'s allows us to fully identify the quantum numbers of the various excitations, including the nature of the twisted vacuum.

Eqs. \eqref{sixtorus}, \eqref{su2u1} and \eqref{parityaction}, together with the modular properties of the $\xi_a$ characters, are all one needs to build the $T^6 /\mathbb{Z}_2^2$ and $T^7/\mathbb{Z}_2^3$ orbifolds and study the allowed discrete torsion phases.

\subsection{The Calabi-Yau case: $T^6/\mathbb{Z}_2^2$}

We start with the construction of the $T^6 /\mathbb{Z}_2^2$ orbifold with generators \eqref{Z2Z2generators}. The untwisted moduli surviving the $\mathbb{Z}_2 \times \mathbb{Z}_2$ projection are the three ($k=1,2,3$) complex structure $U^{(k)} = U_1^{(k)} + i \, U_2^{(k)}$, and  K\"ahler  $T^{(k)} = T_1^{(k)} + i\, T_2^{(k)}$ moduli of $T^6 = T^2\times T^2 \times T^2$, encoded in the block-diagonal metric and $B$ fields,
\begin{equation}
g_{ab}^{(k)} = \sqrt{\frac{T_2^{(k)}}{U_2^{(k)}}}\begin{pmatrix} 1 & U_1^{(k)} \\ U_1^{(k)} & |U^{(k)}|^2 \end{pmatrix}\,, \qquad
B_{ab}^{(k)} = \begin{pmatrix} 0 & T_1^{(k)} \\ -T_1^{(k)} & 0 \end{pmatrix}\,. \label{Z2Z2moduli}
\end{equation}
These moduli represent marginal deformations of the underlying CFT which continuously connect vacua built upon different points in moduli space.
 In particular, the factorised $T^6=T^2\times T^2 \times T^2$ torus with generic values of $U$'s and $T$'s is continuously connected to the lattice \eqref{sixtorus} associated to the $\text{SU} (2)^6$ point, which corresponds to the special values $U^{(k)} = i$ and $T^{(k)}=i$. As a result, without loss of generality, we can focus on the $\text{SU}(2)^6$ point, since the (in)consistency of the orbifold compactification on this special point implies the (in)consistency of orbifold compactifications at a generic point in moduli space. 
 
The partition function in the untwisted sector reads
\begin{equation}
\begin{aligned}
{\mathcal Z} \onech{1}{1} &= \prod_{i=1}^6 \left[ |\xi^i_0 + \xi^i_4|^2 + |\xi^i_2 + \xi^i_{-2}|^2 \right]\,,
\\
{\mathcal Z} \onech{1}{g} &= \prod_{i\in \{1,2,3,4\}} \left[ |\xi^i_0 - \xi^i_4|^2 + |\xi^i_2 - \xi^i_{-2}|^2 \right]\, \prod_{i \in \{5,6\}} \left[ |\xi^i_0 + \xi^i_4|^2 + |\xi^i_2 + \xi^i_{-2}|^2 \right]\,,
\\
{\mathcal Z} \onech{1}{h} &= \prod_{i\in \{3,4,5,6\}} \left[ |\xi^i_0 - \xi^i_4|^2 + |\xi^i_2 - \xi^i_{-2}|^2 \right]\, \prod_{i \in \{1,2\}} \left[ |\xi^i_0 + \xi^i_4|^2 + |\xi^i_2 + \xi^i_{-2}|^2 \right]\,,
\\
{\mathcal Z} \onech{1}{f} &= \prod_{i\in \{1,2,5,6\}} \left[ |\xi^i_0 - \xi^i_4|^2 + |\xi^i_2 - \xi^i_{-2}|^2 \right]\, \prod_{i \in \{3,4\}} \left[ |\xi^i_0 + \xi^i_4|^2 + |\xi^i_2 + \xi^i_{-2}|^2 \right]\,,
\end{aligned}\label{Z2Z2unt}
\end{equation}
where the ${\mathcal Z} \onech{1}{\alpha}$ amplitudes clearly take into account the action \eqref{parityaction} of the generators \eqref{Z2Z2generators}  on the $\xi^i_a$ characters associated to the $i$-th direction. An $S$ modular transformation builds the twisted sectors
\begin{equation}
\begin{aligned}
{\mathcal Z} \onech{g}{1} &= \prod_{i\in \{1,2,3,4\}} \left[ |\xi^i_1 + \xi^i_{-3}|^2 + |\xi^i_{-1} + \xi^i_{3}|^2 \right]\, \prod_{i \in \{5,6\}} \left[ |\xi^i_0 + \xi^i_4|^2 + |\xi^i_2 + \xi^i_{-2}|^2 \right]\,,
\\
{\mathcal Z} \onech{h}{1} &=\prod_{i\in \{3,4,5,6\}} \left[ |\xi^i_1 + \xi^i_{-3}|^2 + |\xi^i_{-1} + \xi^i_{3}|^2 \right]\, \prod_{i \in \{1,2\}} \left[ |\xi^i_0 + \xi^i_4|^2 + |\xi^i_2 + \xi^i_{-2}|^2 \right]\,,
\\
{\mathcal Z} \onech{f}{1} &=\prod_{i\in \{1,2,5,6\}} \left[ |\xi^i_1 + \xi^i_{-3}|^2 + |\xi^i_{-1} + \xi^i_{3}|^2 \right]\, \prod_{i \in \{3,4\}} \left[ |\xi^i_0 + \xi^i_4|^2 + |\xi^i_2 + \xi^i_{-2}|^2 \right]\,,
\end{aligned}\label{Z2Z2twS}
\end{equation}
while a $T$ modular transformation yields
\begin{equation}
\begin{aligned}
{\mathcal Z} \onech{g}{g} &= \prod_{i\in \{1,2,3,4\}} \left[ |\xi^i_1 - \xi^i_{-3}|^2 + |\xi^i_{-1} - \xi^i_{3}|^2 \right]\, \prod_{i \in \{5,6\}} \left[ |\xi^i_0 + \xi^i_4|^2 + |\xi^i_2 + \xi^i_{-2}|^2 \right]\,,
\\
{\mathcal Z} \onech{h}{h} &=\prod_{i\in \{3,4,5,6\}} \left[ |\xi^i_1 - \xi^i_{-3}|^2 + |\xi^i_{-1} - \xi^i_{3}|^2 \right]\, \prod_{i \in \{1,2\}} \left[ |\xi^i_0 + \xi^i_4|^2 + |\xi^i_2 + \xi^i_{-2}|^2 \right]\,,
\\
{\mathcal Z} \onech{f}{f} &=\prod_{i\in \{1,2,5,6\}} \left[ |\xi^i_1 - \xi^i_{-3}|^2 + |\xi^i_{-1} - \xi^i_{3}|^2 \right]\, \prod_{i \in \{3,4\}} \left[ |\xi^i_0 + \xi^i_4|^2 + |\xi^i_2 + \xi^i_{-2}|^2 \right]\,. 
\end{aligned}\label{Z2Z2twT}
\end{equation}
All amplitudes in eqs. \eqref{Z2Z2unt}, \eqref{Z2Z2twS} and \eqref{Z2Z2twT} are uniquely determined by the faithful action of the orbifold group on the untwisted string coordinates and by modular transformations, and represent a sector of the orbifold partition function which is modular invariant \emph{per se}. To complete the orbifold projection in the various twisted sectors one has to add the independent modular orbit associated to the boxes decorated with a blue disc in figure \ref{Z2Z2}. This is the place where the (generalised) discrete torsion appears and affects the projections. To this end, we notice that in each ${\mathcal Z} \onech{\alpha}{1}$ there are four directions which are twisted, and expanding the product along these directions produces $2^4 =16$ terms, each associated to a different fixed $T^2$. As a result, the proposal of \cite{Gaberdiel:2004vx} to associate a different discrete torsion at each fixed point translates into a sign freedom $\epsilon_r$ for each of the sixteen monomials. For the sake of simplicity, and without loss of generality, we consider here a simplified set-up where either all fixed points are treated equally or are split in two groups of eight with and without discrete torsion\footnote{In Appendix \ref{AppA} we consider the most generic case with sixteen independent signs.}. We thus write 
\begin{equation}
\begin{aligned}
{\mathcal Z} \onech{g}{h} &= \epsilon \prod_{i\in \{1,2\}} \left[ |\xi^i_1 + \xi^i_{-3}|^2 + \eta_{i} |\xi^i_{-1} + \xi^i_{3}|^2 \right]
\prod_{i\in \{3,4\}} \left[  |\xi^i_1 - \xi^i_{-3}|^2 + \eta_{i} |\xi^i_{-1} - \xi^i_{3}|^2 \right]\,
\\
&\quad\times
\prod_{i \in \{5,6\}} \left[ |\xi^i_0 - \xi^i_4|^2 + |\xi^i_2 - \xi^i_{-2}|^2 \right]\,.
\end{aligned}\label{Z2Z2gh}
\end{equation}
Here the overall sign $\epsilon$ is related to the discrete torsion of Vafa \cite{Vafa:1986wx}, while the relative signs $\eta_i$ realise the generalised discrete torsion of \cite{Gaberdiel:2004vx}. 
 A $T$ modular transformation yields
\begin{equation}
\begin{aligned}
{\mathcal Z} \onech{g}{f} &= \epsilon \prod_{i\in \{1,2\}} \left[ |\xi^i_1 - \xi^i_{-3}|^2 + \eta_{i} |\xi^i_{-1} - \xi^i_{3}|^2 \right]\, 
\prod_{i\in \{3,4\}} \left[ |\xi^i_1 + \xi^i_{-3}|^2 + \eta_{i} |\xi^i_{-1} + \xi^i_{3}|^2 \right]\,
\\
&\quad\times
\prod_{i \in \{5,6\}} \left[ |\xi^i_0 - \xi^i_4|^2 + |\xi^i_2 - \xi^i_{-2}|^2 \right]\,,
\end{aligned}\label{Z2Z2gf}
\end{equation}
and completes the projection.

To determine the contribution of the $g$-twisted sector to the Hodge numbers of the $\text{CY}_3$ manifold, one should focus on the combinations of characters with smallest conformal weight, which can only involve $\xi_0$ and $\xi_{\pm 1}$. There are exactly sixteen such terms which are in one-to-one correspondence with the sixteen fixed tori. Each combination determines the ground state of the Hilbert space ${\mathcal H}_g^r$ attached to the $r$-th fixed locus. Let us consider, for instance, the combination of characters
\begin{equation}
|\xi_1^1\, \xi_{-1}^2\, \xi_{-1}^3\, \xi_1^4\, \xi_0^5\, \xi_0^6|^2 \,,
\end{equation}
of conformal weight $1/2$. This corresponds to the twisted ground state
\begin{equation}
|\Omega_g^r\rangle= \sigma^1_{a} \tilde\sigma^1_{a} \sigma^2_{\dot a} \tilde\sigma^2_{\dot a} \sigma^3_{\dot a} \tilde\sigma^3_{\dot a} \sigma^4_{a} \tilde\sigma^4_{a} |0,\tilde 0\rangle \,,
\end{equation}
where we have mapped $\xi_{1} \to \sigma_a$ and $\xi_{-1} \to \sigma_{\dot a}$, and similarly for the anti-holomorphic sector. This identification matches the $\text{U} (1)$ charges and conformal weight of characters and twist fields.  The ground state $|\Omega_g^r\rangle$ is a two-form representing the class of exceptional divisor obtained upon resolving the $r$-th singularity.  From the amplitudes ${\mathcal Z} \onech{g}{\alpha}$ we read that this two-form is invariant under the action of $g$ while picks up the sign $\epsilon \eta_2 \eta_3$ under both $h$ and $f$.

To determine the contribution of each $g$-twisted fixed locus to the Hodge numbers, one has to dress this bosonic ground state with a suitable number of RR zero-modes from the fermionic coordinates. The complex combinations $\psi_\pm^i = (\psi_0^i \pm i \tilde \psi_0^i)/2$ of the R zero-modes satisfy the algebra of fermionic harmonic oscillators
\begin{equation}
\{ \psi_\pm ^i , \psi_\mp^j \} = \delta^{ij}\,, \qquad \{ \psi_\pm ^i , \psi_\pm^j \} = 0\,,
\end{equation}
which identifies $\psi_+^i$ and $\psi_-^i$ as the creation and annihilation operators, respectively. Therefore, one finds the invariant combinations
\begin{equation}
|\Omega_g^r\rangle \qquad \text{and}\qquad \psi^5_+ \psi^6_+ |\Omega_g^r \rangle\,,
\end{equation}
if $\epsilon \eta_2 \eta_3 = +1$, or
\begin{equation}
\psi^5_+ |\Omega_g^r\rangle \qquad \text{and}\qquad \psi^6_+ |\Omega_g^r \rangle\,,
\end{equation}
if $\epsilon \eta_2 \eta_3= -1$. As a result, a fixed point without discrete torsion ($\epsilon \eta_2 \eta_3 = +1$) contributes with one unit to $h^{1,1}$ and $h^{2,2}$, while a fixed point supporting a non-trivial discrete torsion ($\epsilon \eta_2 \eta_3 = -1$) contributes with one unit to $h^{1,2}$ and $h^{2,1}$.

To determine the full structure of the $h$ and $f$-twisted sectors, one needs to build ${\mathcal Z}\onech{h}{g}$ and ${\mathcal Z}\onech{f}{g}$. These can be obtained from eqs. \eqref{Z2Z2gh} and \eqref{Z2Z2gf} via an $S$ modular transformation. The result clearly depends on the choice of the signs $\eta_i$. In fact, using the modular properties of the $\xi_a$ characters, it is straightforward to show that
\begin{equation}
\begin{aligned}
|\xi_1 + \xi_{-3}|^2 + |\xi_{-1} + \xi_{3}|^2 &\to |\xi_0 - \xi_{4}|^2 + |\xi_{2} - \xi_{-2}|^2\,,
\\
|\xi_1 - \xi_{-3}|^2 + |\xi_{-1} - \xi_{3}|^2 &\to |\xi_1 - \xi_{-3}|^2 + |\xi_{-1} - \xi_{3}|^2\,,
\end{aligned}
\end{equation}
while
\begin{equation}
\begin{aligned}
|\xi_1 + \xi_{-3}|^2 - |\xi_{-1} + \xi_{3}|^2 &\to -i \left[ (\xi_0 -\xi_4 )(\bar \xi_2 -\bar\xi_{-2}) - (\xi_2 - \xi_{-2})  (\bar\xi_0 -\bar\xi_4 )\right]\,,
\\
|\xi_1 - \xi_{-3}|^2 - |\xi_{-1} - \xi_{3}|^2 &\to +i \left[ (\xi_1 -\xi_{-3} )(\bar \xi_{-1} -\bar\xi_{3}) - (\xi_{-1} - \xi_{3})  (\bar\xi_{1} -\bar\xi_{-3} )\right] \,.
\end{aligned}\label{dtSt}
\end{equation}
It is clear that only the choice $\eta_i  = +1$, for every $i=1,\ldots , 4$, is allowed, since only in this case ${\mathcal Z}\onech{h}{g}$ (${\mathcal Z}\onech{f}{g}$) contains the very same states present in ${\mathcal Z}\onech{h}{1}$ (${\mathcal Z}\onech{f}{1}$), thus providing a consistent operator interpretation, and thus a consistent CFT. Indeed, from ${\mathcal Z}\onech{h}{1}$ we read that, along any twisted direction, left-handed and right-handed twist fields carry the same charge, for instance
\begin{equation}
{\mathcal Z} \onech{h}{1} \quad \Rightarrow \quad \sigma^3_{a} \tilde\sigma^3_{a} \sigma^4_{\dot a} \tilde\sigma^4_{\dot a} \sigma^5_{\dot a} \tilde\sigma^5_{\dot a} \sigma^6_{a} \tilde\sigma^6_{a} |0,\tilde 0\rangle \,.\label{htvacuum}
\end{equation}
This state is present (modulo signs) in all amplitudes ${\mathcal Z} \onech{h}{\alpha}$ if $\eta_i =+1$, thus providing a consistent CFT. On the contrary,
for the choice $\eta_i = -1$,  the transformations \eqref{dtSt} imply that ${\mathcal Z}\onech{h}{g}$ would project a twisted vacuum built upon left-handed and right-handed twist fields of opposite charge. For instance, for $\eta_2 =-1$, one finds
\begin{equation}
{\mathcal Z} \onech{h}{g} \quad \Rightarrow \quad \sigma^3_{a} \tilde\sigma^3_{ a} \sigma^4_{\dot a} \tilde\sigma^4_{ a} \sigma^5_{\dot a} \tilde\sigma^5_{\dot a} \sigma^6_{a} \tilde\sigma^6_{a} |0,\tilde 0\rangle,
\end{equation}
which involves $\sigma^4_{\dot a}$ and $\tilde\sigma^4_{a}$ and is incompatible with the structure of the $h$-twisted vacuum \eqref{htvacuum}, one reads from ${\mathcal Z}\onech{h}{1}$.

Clearly, similar arguments hold for the $f$-twisted sector, and for more general discrete torsions involving sixteen independent signs, as shown in the appendix \ref{AppA}. As a result, all fixed points, in every twisted sector, \emph{must} support the same discrete torsion. To iterate, a consistent string compactification on the $T^6/\mathbb{Z}_2^2$ orbifold only admits the canonical discrete torsion of \cite{Vafa:1986wx}, while the generalised discrete torsion of \cite{Gaberdiel:2004vx} does not yield a consistent CFT\footnote{Of course, other discrete deformations are allowed as classified by the equivariant cohomology groups \cite{BoyleSmith:2026oay}. As already stressed several times, here we only consider the topological deformation associated to the discrete torsion of \cite{Vafa:1986wx} and its (im)possible generalisation \emph{\`a la} \cite{Gaberdiel:2004vx}.}. This conclusion can be straightforwardly generalised to $\mathbb{Z}_N \times \mathbb{Z}_M$ orbifolds.

\subsubsection{Modular invariant partition functions at the rational point}

In support to the analysis discussed above, and thus to the impossibility to turn on generalised discrete torsion, we have thoroughly analysed the various modular invariant partition functions at rational points. For a RCFT, with characters $\chi_a$, the one-loop partition function reads
\begin{equation}
{\mathcal Z} = \sum_{a,b} \bar \chi_a \, X_{ab} \, \chi_b \,,
\end{equation}
where the matrix $X_{ab}$ encodes the orbifold projection, and is bound to satisfy the modular invariant constraints
\begin{equation}
T^\dagger \, X \, T = X \,, \qquad S^\dagger \, X \, S = X\,. \label{eq:modinvcond}
\end{equation}
Different choices of generalised discrete torsions are then associated to different solutions.

To this end, we have considered the compactification of the type IIB superstring on the $T^6 /\mathbb{Z}_2^2$ orbifold, where the six-torus is at the rational point $\text{SO} (4)^3$. The RCFT consists of 1024 characters, whose identity is
\begin{equation}
\begin{aligned}
\chi_1 &= (vooo + o vvv -scsc-cscs) (oooooo+vvvvvv) 
\\
&\quad + (ovoo+vovv-sccs-cssc)(vvoooo+oovvvv)
\\
&\quad + (ooov+vvvo-sscc-ccss)(vvvvoo+oooovv)
\\
&\quad + (oovo+vvov-ssss-cccc)(oovvoo+vvoovv)\,,
\end{aligned}
\end{equation}
written in terms of SO(2) characters. The contributions to the first parenthesis are associated to the RNS fermions, while those in the second set are associated to the compact world-sheet bosons. The characters have a built-in GSO projection which respects space-time supersymmetry. The untwisted sector comprises three massless real characters, aside from the identity above, while the $g$, $h$ and $f$ twisted sectors comprise 16 massless pairs of complex conjugate characters, $\chi$ and $\chi^c$, each.

In a geometric compactification, the modular invariant partition function does not mix the untwisted and twisted sectors. In this case, the three extra untwisted real massless characters  always appear diagonally in ${\mathcal Z}$, and thus yield a vector and an hypermultiplet, each, contributing 3 units to the Hodge numbers $h^{1,2}$ and $h^{1,1}$, as expected. The contributions from the twisted characters to the cohomology of the resulting Calabi-Yau space depends on how the complex pairs enter the partition function. In particular, the combination $\bar\chi \, \chi^c + \bar\chi^c \, \chi$ yields a massless hypermultiplet, and thus contributes one unit to $h^{1,1}$, while the diagonal combination $|\chi|^2 + |\chi^c |^2$ yields a vector multiplet and thus contributes one unit to $h^{1,2}$. The canonical compactification, without discrete torsion, is then associated to the diagonal modular invariant partition function, while the discrete torsion of \cite{Vafa:1986wx} corresponds to the (full) charge-conjugation modular invariant. The generalised discrete torsion of \cite{Gaberdiel:2004vx} would then correspond to a modular invariant $X$ which, in a given twisted sector, combines some massless characters diagonally and others with their conjugate. A thorough study of the  conditions \eqref{eq:modinvcond} shows that it is impossible to build such \emph{mixed} modular invariants, and therefore the generalised discrete torsion is not compatible with one-loop modular invariance at the rational point, where particle interpretation is built-in since characters are in one-to-one correspondence with chiral primary operators of the SCFT.

\subsection{The $G_2$ manifold case: $T^7/\mathbb{Z}_2^3$} \label{ssec:G2orbifoldgenus1}

The analysis of the $G_2$ manifold is similar to that of the previous case. Again, we start from seven copies of a compact boson with radius $R=\sqrt{\alpha'}$, and twist them by the $\mathbb{Z}_2^3$ orbifold generated by \eqref{G2gen}. Also in this case, reflections and shifts break $\text{SU} (2) \to \text{U}(1)_4$, and act as eq. \eqref{parityaction} and
\begin{equation}
\text{shift}: \qquad |\xi_0 + \xi_4 |^2 + |\xi_2 + \xi_{-2}|^2 \to |\xi_0 - \xi_4 |^2 - |\xi_2 - \xi_{-2}|^2 \,,
\end{equation}
on a given compact coordinate. Using this information, one can build the full partition function. However, to keep the presentation light, we shall focus on the $c$ and $ab$ twisted sectors which involve the independent orbit supporting discrete torsion, see figure \ref{G2}. We start from 
\begin{equation}
\begin{aligned}
{\mathcal Z}\onech{1}{c} &= \prod_{i\in \{1,3,5,7\}}\left[|\xi^i_0 - \xi^i_4 |^2 + |\xi^i_2 - \xi^i_{-2}|^2 \right] \prod_{i\in \{2,4,6\}} 
\left[|\xi^i_0 + \xi^i_4 |^2 + |\xi^i_2 + \xi^i_{-2}|^2 \right] \,,
\\
{\mathcal Z}\onech{c}{1} &= \prod_{i\in \{1,3,5,7\}}\left[|\xi^i_1 + \xi^i_{-3} |^2 + |\xi^i_{-1} + \xi^i_{3}|^2 \right] \prod_{i\in \{2,4,6\}} 
\left[|\xi^i_0 + \xi^i_4 |^2 + |\xi^i_2 + \xi^i_{-2}|^2 \right] \,,
\\
{\mathcal Z}\onech{c}{c} &= \prod_{i\in \{1,3,5,7\}}\left[|\xi^i_1 - \xi^i_{-3} |^2 + |\xi^i_{-1} - \xi^i_{3}|^2 \right] \prod_{i\in \{2,4,6\}} 
\left[|\xi^i_0 + \xi^i_4 |^2 + |\xi^i_2 + \xi^i_{-2}|^2 \right] \,,
\end{aligned}
\end{equation}
which, as usual, are uniquely fixed by the faithful  action of $c$ on the untwisted sector and by $S$ and $T$ modular transformations. The independent modular orbit, which completes the projection on the $c$-twisted sector, includes the amplitudes
\begin{equation}
\begin{aligned}
{\mathcal Z}\onech{c}{ab} &= \epsilon \left[|\xi^2_0 - \xi^2_4 |^2 - |\xi^2_2 - \xi^2_{-2}|^2 \right]  \prod_{i\in{\{1,7\}}}\left[|\xi^i_1 + \xi^i_{-3} |^2 + \eta_{i} |\xi^i_{-1} + \xi^i_{3}|^2 \right] 
\\
&\quad \times \prod_{i\in \{3,5\}} \left[ |\xi^i_1 - \xi^i_{-3} |^2 +\eta_{i} |\xi^i_{-1} - \xi^i_{3}|^2 \right] 
\prod_{i\in \{4,6\}} 
\left[|\xi^i_0 - \xi^i_4 |^2 + |\xi^i_2 - \xi^i_{-2}|^2 \right]  \,,
\\
{\mathcal Z}\onech{c}{abc} &= \epsilon\left[|\xi^2_0 - \xi^2_4 |^2 - |\xi^2_2 - \xi^2_{-2}|^2 \right]  \prod_{i\in{\{1,7\}}}\left[ |\xi^i_1 - \xi^i_{-3} |^2 + \eta_{i} |\xi^i_{-1} - \xi^i_{3}|^2 \right] 
\\
&\quad \times \prod_{i\in \{3,5\}} \left[  |\xi^i_1 + \xi^i_{-3} |^2 +\eta_{i} |\xi^i_{-1} + \xi^i_{3}|^2 \right] 
\prod_{i\in \{4,6\}} \left[|\xi^i_0 - \xi^i_4 |^2 + |\xi^i_2 - \xi^i_{-2}|^2 \right]  \,,
\end{aligned}
\end{equation}
which are related by a $T$ transformation, ${\mathcal Z}\onech{c}{abc} = T\,{\mathcal Z}\onech{c}{ab}$. As in the $T^6 /\mathbb{Z}_2^2$ case, we have added the overall sign $\epsilon$, encoding the standard discrete torsion, and the relative signs $\eta_i$ for the twisted directions, encoding the generalised discrete torsion. Also in this case, $\eta_i =+1$, for any $i$, treats all fixed points on equal footing, while if $\eta_i =-1$, for at least one value of $i$, eight fixed points support generalised discrete torsion and eight do not. Of course, this is not the most general situation, which would involve sixteen independent signs, but it is sufficient for our purposes. Following the analysis of \cite{Gaberdiel:2004vx}, the $c$ twisted sector yields the Betti numbers $(b^2 , b^3 ) =  (8,8)$ if $\epsilon=+1$ and $\eta_i = +1$ or $(b^2 , b^3 ) =  (0,16)$ if $\epsilon=-1$ and $\eta_i = +1$, while one gets $(b^2 , b^3 ) =  (4,12)$ if $\eta_i = -1$, for any choice of $\epsilon$. Adding the contribution of the untwisted and $a$ and $b$ twisted sectors, and allowing for the most general discrete torsion, one would get \cite{Gaberdiel:2004vx} the family of $G_2$ manifolds with Betti numbers
\begin{equation}
(b^0 , \ldots , b^7) = (1, 0 , 8 +\ell , 47-\ell, 47-\ell, 8+\ell , 0 ,1)\,, \qquad \text{for}\quad \ell = 0,\ldots , 8\,.
\end{equation}

Although, the cohomology of the $G_2$ manifold is fully encoded in the massless sectors, which are naively consistent for any choice of $\ell$, to build a \emph{bona fide} CFT one should  verify that also the massive sectors provide a consistent operator interpretation. The $ac$ and $bc$ twisted sectors are  fully determined by the ${\mathcal Z}\onech{1}{ac}$ and ${\mathcal Z}\onech{1}{bc}$ untwisted amplitudes, see figure \ref{G2}, and therefore do not bring in any surprise. On the contrary, the $ab$ and $abc$ sectors need more care since they involve the additional projection associated to the independent modular orbit, which depends on the choice of generalised discrete torsion. Let us focus on 
\begin{equation}
\begin{aligned}
{\mathcal Z}\onech{ab}{1} &=  \left[(\xi^2_1 + \xi^2_{-3})(\bar\xi^2_{-1} + \bar\xi^2_{3}) + (\xi^2_{-1} + \xi^2_{3}) (\bar\xi^2_1 + \bar\xi^2_{-3}) \right]\prod_{i\in \{1,7\}} \left[|\xi^i_0 + \xi^i_4 |^2 + |\xi^i_2 + \xi^i_{-2}|^2 \right]
\\
&\quad \times
\prod_{i\in \{3,4,5,6\}}\left[|\xi^i_1 + \xi^i_{-3} |^2 + |\xi^i_{-1} + \xi^i_{3}|^2 \right]  \,,
\end{aligned}\label{G2abt}
\end{equation}
which is obtained as an $S$ transformation of
\begin{equation}
\begin{aligned}
{\mathcal Z}\onech{1}{ab} &=  \left[|\xi^2_0 - \xi^2_4 |^2 - |\xi^2_2 - \xi^2_{-2}|^2 \right] 
\prod_{i\in \{1,7\}} \left[|\xi^i_0 + \xi^i_4 |^2 + |\xi^i_2 + \xi^i_{-2}|^2 \right]
\\
&\quad \times
\prod_{i\in \{3,4,5,6\}}\left[|\xi^i_0 - \xi^i_4 |^2 + |\xi^i_2 - \xi^i_{-2}|^2 \right]  \,.
\end{aligned}
\end{equation}
From \eqref{G2abt} we read that the coordinates 1 and 7 are untwisted and involve the standard Narain lattice, the coordinate 2 is also untwisted but now with shifted windings, while the coordinates $3,\ldots , 6$ are twisted, consistently with the action of 
\begin{equation}
ab = (+,+,-,-,-,-,+|0,\nicefrac{1}{2},0,0,0,0,0)\,.
\end{equation}
As in the $\text{CY}_3$ case, the twisted vacuum involves left-handed and right-handed twist fields with equal U(1) charge, although now it also carries a non-trivial winding number.

The action of the $c$ element on this twisted spectrum is obtained by ${\mathcal Z}\onech{c}{ab}$ via an $S$ modular transformation. For all $\eta_i=+1$, one gets
\begin{equation}
\begin{aligned}
{\mathcal Z}\onech{ab}{c} &= \epsilon \left[(\xi^2_1 + \xi^2_{-3})(\bar\xi^2_{-1} + \bar\xi^2_{3}) + (\xi^2_{-1} + \xi^2_{3}) (\bar\xi^2_1 + \bar\xi^2_{-3}) \right]  \prod_{i\in{\{1,7\}}}\left[|\xi^i_0 - \xi^i_4 |^2 + |\xi^i_{2} - \xi^i_{-2}|^2 \right] 
\\
&\quad \times \prod_{i\in \{3,5\}} \left[|\xi^i_1 - \xi^i_{-3} |^2 + |\xi^i_{-1} - \xi^i_{3}|^2 \right] 
\prod_{i\in \{4,6\}} 
\left[|\xi^i_1 + \xi^i_{-3} |^2 + |\xi^i_{-1} + \xi^i_{3}|^2 \right] 
\,,
\end{aligned}
\end{equation}
which is clearly compatible with ${\mathcal Z}\onech{ab}{1}$ since the two amplitudes contain the same states (modulo signs). On the contrary, turning on generalised discrete torsion, which amounts to choosing at least one $\eta_i$ negative, violates the operator interpretation of the partition function. For instance, for $\eta_3=-1$ one gets
\begin{equation}
\begin{aligned}
{\mathcal Z}\onech{ab}{c} &= \left[(\xi^2_1 + \xi^2_{-3})(\bar\xi^2_{-1} + \bar\xi^2_{3}) + (\xi^2_{-1} + \xi^2_{3}) (\bar\xi^2_1 + \bar\xi^2_{-3}) \right]  \prod_{i\in{\{1,7\}}}\left[|\xi^i_0 - \xi^i_4 |^2 + |\xi^i_{2} - \xi^i_{-2}|^2 \right] 
\\
&\quad \times 
i \left[(\xi^3_1 - \xi^3_{-3})(\bar\xi^3_{-1}- \bar\xi^3_{3}) - (\xi^3_{-1} - \xi^3_{3}) (\bar\xi^3_1 - \bar\xi^3_{-3}) \right]
\\
&\quad \times
\left[|\xi^5_1 - \xi^5_{-3} |^2 + |\xi^5_{-1} - \xi^5_{3}|^2 \right] 
\prod_{i\in \{4,6\}} 
\left[|\xi^i_1 + \xi^i_{-3} |^2 + |\xi^i_{-1} + \xi^i_{3}|^2 \right]  \,,
\end{aligned}
\end{equation}
clearly incompatible with ${\mathcal Z}\onech{ab}{1}$.

A similar analysis holds for the $abc$ twisted sector, and thus we are led to the conclusion that only the discrete torsion of Vafa \cite{Vafa:1986wx} yields a consistent CFT\footnote{Here, the same comment of footnote 3 applies.}, and the $T^6/\mathbb{Z}_2^3$ orbifold with generators \eqref{G2gen}, can only build the two $G_2$ manifolds with Betti numbers
\begin{equation}
\begin{aligned}
(b^0 ,\ldots , b^7) &= (1, 0 , 8 , 47 , 47, 8  , 0 ,1)\,, \qquad &\text{for} \quad \epsilon=+1\,,
\\
(b^0 , \ldots , b^7) &= (1, 0 , 16 , 39, 39, 16 , 0 ,1)\,, \qquad &\text{for} \quad \epsilon=-1\,.
\end{aligned}
\end{equation}

\section{Modular invariance and factorisation at genus 2}\label{sec:genus2}

We have seen how a consistent operator interpretation of the one-loop vacuum energy implies that only the canonical discrete torsion is acceptable, while any attempt to assign different signs at different fixed points, although compatible with modular invariance, results into an inconsistent particle interpretation. This is manifest at rational points in moduli space as, for instance, the  $\text{SU}(2)^6$ and the $\text{SU}(2)^7$ points, where the partition function allows one to extract information about the various quantum numbers of the string excitations.

For  rational CFT's, it has been shown \cite{Antoniadis:1986rn, Moore:1988qv,  Gaberdiel:2010jf} that a consistent operator interpretation of the one-loop amplitude is closely intertwined with genus two (and higher) modular invariance. Although it is reasonable to believe that this connection still holds at a generic point of moduli space, no direct proof exists. For this reason, we shall now turn to study the constraints on the possible generalised discrete torsions, arising from modular invariance and factorisation of genus two amplitudes. We will concentrate on the $T^6/\mathbb{Z}_2^2$ case, but similar results are expected to hold also for the $T^7/\mathbb{Z}_2^3$ orbifold. Since we shall study the genus-two amplitude at a generic point of moduli space, this analysis will provide a complementary proof of the results obtained at one loop.

\begin{figure}
\centering
\includegraphics[scale=0.15]{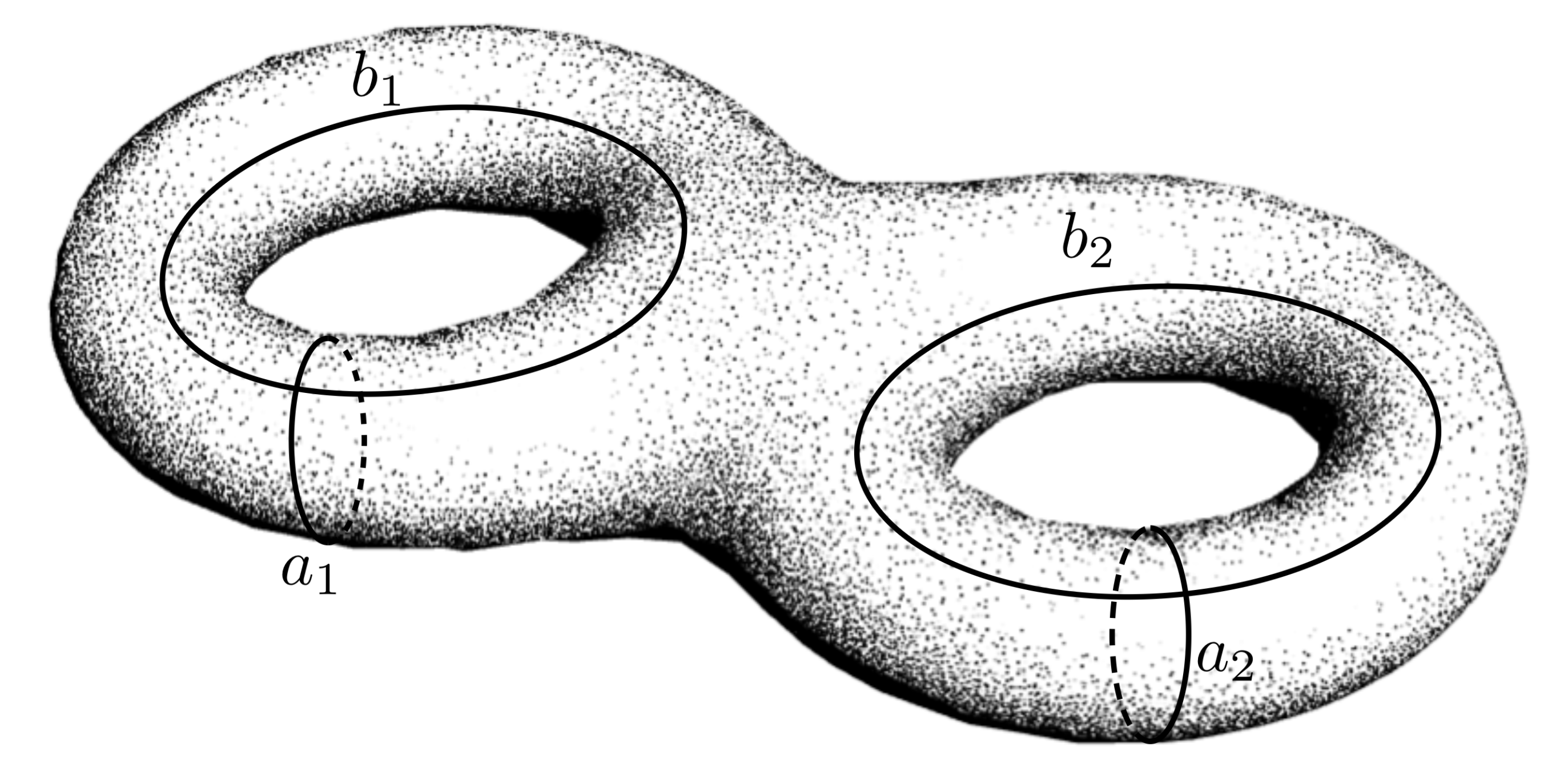}
\caption{The genus two Riemann surface with the canonical $a$ and $b$ cycles.}
\label{twotorus}
\end{figure}

A genus two Riemann surface $\Sigma_2$, with its canonical $a$ and $b$ cycles is represented in figure \ref{twotorus}. Its shape and size are encoded in the period matrix $\Omega_{ij}$ defined by the relations
\begin{equation}
\oint_{a_i} \omega_j = \delta_{ij}\,, \qquad \oint_{b_i} \omega_j = \Omega_{ij}\,,
\end{equation}
with $\omega_j$ a basis of holomorphic one-forms. The set of modular transformations which connect equivalent $\Sigma_2$'s generates $\text{Sp} (4;\mathbb{Z})$, and acts on the period matrix as
\begin{equation}
\Omega \to (A \,\Omega + B)\, (C \, \Omega + D)^{-1}\,, \qquad 
\begin{pmatrix} A & B \\ C & D\end{pmatrix} \in \text{Sp} (4; \mathbb{Z})\,.
\end{equation}

A closed string coordinate $X (z , \bar z )$ on $\Sigma_2$ is defined by its periodicity along the $a$ and $b$ cycles. These are encoded in the twists 
$\zeta = \twoch{\zeta^{a_1}}{\zeta^{a_2}}{\zeta^{b_1}}{\zeta^{b_2}}$. In the case of  periodic or antiperiodic boundary conditions one has to deal with the sixteen choices
\begin{equation}
\begin{aligned}
&\zeta_0= \twoch{0}{0}{0}{0} \,, 
\\  
&\zeta_1=\twoch{0}{0}{0}{\nicefrac{1}{2}}  \,, 
\\  
&\zeta_2= \twoch{\nicefrac{1}{2}}{\nicefrac{1}{2}}{0}{\nicefrac{1}{2}} \,, 
\\  
&\zeta_3= \twoch{0}{\nicefrac{1}{2}}{0}{0} \,, 
\end{aligned} 
    \qquad 
\begin{aligned}
&\zeta_4= \twoch{0}{\nicefrac{1}{2}}{0}{\nicefrac{1}{2}} \,, 
\\  
&\zeta_5=\twoch{0}{\nicefrac{1}{2}}{\nicefrac{1}{2}}{\nicefrac{1}{2}}  \,, 
\\  
&\zeta_6=\twoch{\nicefrac{1}{2}}{\nicefrac{1}{2}}{\nicefrac{1}{2}}{\nicefrac{1}{2}}  \,, 
\\  
&\zeta_7=\twoch{\nicefrac{1}{2}}{\nicefrac{1}{2}}{0}{0} \,, 
\end{aligned} 
    \qquad 
\begin{aligned}
&\zeta_8=\twoch{0}{\nicefrac{1}{2}}{\nicefrac{1}{2}}{0} \,, 
\\  
& \zeta_{9}=\twoch{\nicefrac{1}{2}}{\nicefrac{1}{2}}{\nicefrac{1}{2}}{0}\,, 
\\  
&\zeta_{10}=\twoch{\nicefrac{1}{2}}{0}{0}{0} \,, 
\\  
&\zeta_{11}=\twoch{0}{0}{\nicefrac{1}{2}}{0} \,, 
\end{aligned} 
    \qquad 
\begin{aligned}
&\zeta_{12}=\twoch{\nicefrac{1}{2}}{0}{\nicefrac{1}{2}}{0}  \,, 
\\  
&\zeta_{13}=\twoch{\nicefrac{1}{2}}{0}{\nicefrac{1}{2}}{\nicefrac{1}{2}} \,, 
\\  
&\zeta_{14}=\twoch{0}{0}{\nicefrac{1}{2}}{\nicefrac{1}{2}} \,, 
\\  
&\zeta_{15}=\twoch{\nicefrac{1}{2}}{0}{0}{\nicefrac{1}{2}} \,. 
\end{aligned}\label{eq:twists}
\end{equation}

\subsection{The Calabi-Yau case: $T^6/\mathbb{Z}_2^2$ orbifold}

We can now move to discuss the genus-two vacuum energy for the $T^6/\mathbb{Z}_2^2$ orbifold, with generators \eqref{Z2Z2generators}. Also in this case, we shall focus on the world-sheet bosons only, since this will suffice to study the (in)consistency of the generalised discrete torsion.

Denoting by $T [\zeta ]$ the contribution of a single real boson twisted by one of the $\zeta$'s in eq. \eqref{eq:twists}, the two-loop partition function can be schematically written as
\begin{equation}
\label{eq:genus2}
    \mathcal{Z}_2= \frac{1}{16}\sum_{i,j=0}^{15} \big (T[\zeta_j] \big )^2 \big (T [\zeta_i+\zeta_j] \big )^2 \big (T [\zeta_i] \big )^2\,.
\end{equation}
The ordering of the $T$'s in eq. \eqref{eq:genus2} reflects the action of the elements of the orbifold group. For instance, the pairs $\zeta_{j=0}$ and $\zeta_{i\neq 0}$ are associated to $h$, the pairs $\zeta_{j\neq 0}$ and $\zeta_{i=0}$ are associated to $g$, while $\zeta_{i\neq 0}$ and $\zeta_{j=i}$ are associated to $f$. The remaining terms encode a mixed action of $g$, $h$ and $f$.

To derive the explicit expressions of the $T$ amplitudes, we start observing that the trivial twist $\zeta_0$ describes a compact boson which is periodic along all $a$ and $b$ cycles of $\Sigma_2$. As a result,
\begin{equation}
    T[\zeta_0]= \Lambda_{1,1} (\Omega, \bar \Omega) \,,
\end{equation}
with the genus-two Narain lattice \cite{Lerche:1988np}
\begin{equation}
\Lambda_{1,1}= \sum_{m,w \in \mathbb{Z}^2} \frac{e^{i \pi \left(  p_\text{L}^t \, \Omega \, p_\text{L} -  p_\text{R}^t \, \bar{\Omega}\, p_\text{R} \right)}}{{\mathcal Z}_0(\Omega) \, \bar{\mathcal Z}_0(\bar \Omega)}
\,,
 \qquad  p_{\text{L,R}}  = \sqrt{\frac{\alpha'}{2}} \left( \frac{m}{R} \pm \frac{w R}{\alpha '} \right)\,. \label{eq:g2narain}
\end{equation}
The sum over the integers $m,w$ encodes the classical contribution, while ${\mathcal Z}_0(\Omega) \, \bar{\mathcal Z} _0(\bar \Omega) = \text{det}' (- \Delta )$ is associated to the quantum fluctuations. The explicit expression of ${\mathcal Z}_0(\Omega)$ is quite cumbersome, but fortunately we only need to know that ${\mathcal Z}_0(\Omega) \, \bar{\mathcal Z} _0(\bar \Omega) $ has modular weight $(-\nicefrac{1}{2} , -\nicefrac{1}{2})$ and, in the factorisation limit, reduces to a product of Dedekind eta functions \cite{Verlinde:1986kw}.

\begin{figure}
\centering
\includegraphics[scale=0.25]{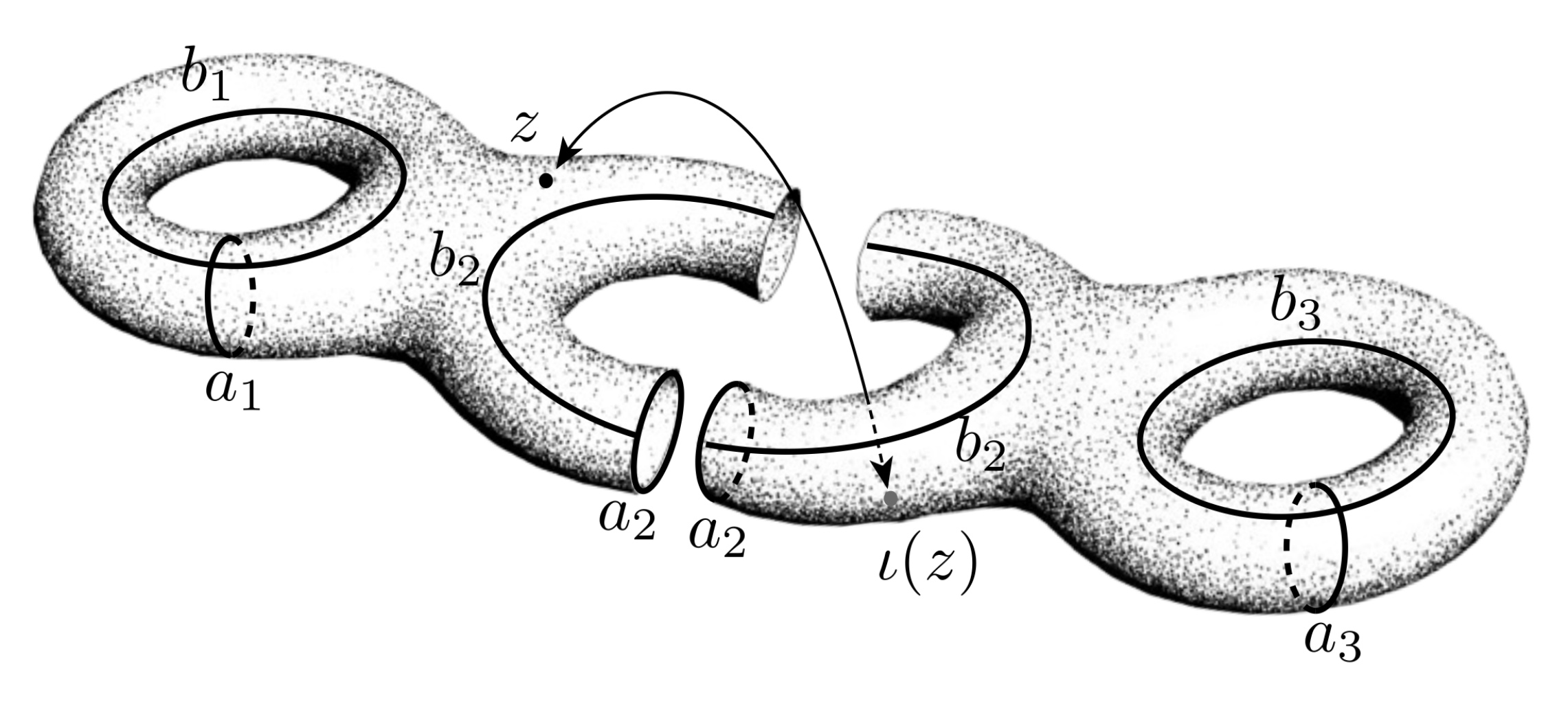}
\caption{The double cover genus three surface for the twist $\zeta_1$. For a boson antiperiodic along $b_2$, cut the Riemann surface $\Sigma_2$ following the $a_2$ cycle. Double it and connect the points on the two copy with the involution $\iota$ reflecting the points across $a_2$. The resulting Riemann surface has genus three, with cycles $a_1$, $b_1$, $a_3 = \iota (a_1)$, $b_3 = \iota (b_1)$, $a_2$ and $2 b_2$.  }
\label{doublecover}
\end{figure}

The amplitudes $ T[\zeta_{i\neq 0}]$ are more subtle, since they are associated to anti-periodic bosons. A natural way to deal with this case is to double cover $\Sigma_2$ so that the boson becomes periodic along all cycles. As an example, let us consider the twist $\zeta_1$. The compact boson is now anti-periodic, \emph{i.e.} double valued, along the $b_2$ cycle. To make it single valued, cut the original surface $\Sigma_2$ along the $a_2$ cycle, the one which intersects $b_2$. Double the surface and glue the two copies along the common boundary $a_2$, as depicted in figure \ref{doublecover}. To preserve the orientation, the points on the two copies are identified via the anti-holomorphic involution $\iota$. The resulting Riemann surface $\Sigma_3$ has genus three\footnote{Notice that such a procedure is not restricted to genus two, but can be extended to any genus-$g$ Riemann surface, resulting in a surface of genus $2g-1$.}, and one-cycles $a_1$, $a_2$, $a_3 = \iota (a_1)$, $b_1$, $2 b_2$ and $b_3 = \iota (b_1)$. As a result, the compact boson is now periodic along all cycles, with the identification
\begin{equation}
X(\iota(z),\overline{\iota(z)})=-X(z,\bar{z}) \quad \text{mod} \quad 2\pi  R\,.
\end{equation}
One can then introduce a holomorphic one-form $\nu$, called the \emph{Prym differential}, which is odd under the involution $\iota$ and normalised such that
\begin{equation}
\oint_{a_1} \nu = - \oint_{a_3} \nu = 1\,, \qquad 
\oint_{b_1} \nu = - \oint_{b_3} \nu = \pi_1 \,, \qquad 
\oint_{a_2} \nu = 0= \oint_{2b_2} \nu \,. 
\end{equation}
The period $\pi_1$, associated to the twist $\zeta_1$, defines a torus ($z\sim z+m+n\pi_1$) known as the \emph{Prym variety}  \cite{fay1973theta}, and induces new features which have no counterpart at genus one: the Laplace equation with twisted boundary conditions admits zero modes living on the Prym variety. Intuitively, in the case at hand, the existence of these zero modes is related to the fact that the twisted boundary conditions do not concern the $a_1$ and $b_1$ cycles, which can then support them\footnote{See Appendix \ref{twoloopc1equiv} for more details.}.

The treatment of the quantum fluctuations associated to a non-trivial twist $\zeta_i$ is, in general, more involved. However, for this case with periodic/antiperiodic boundary conditions, one can rely on the intersection of the two branches for a $c=1$ system \cite{Dijkgraaf:1987vp, Ginsparg:1987eb}
\begin{equation}
\mathcal{Z}_g^{S^1}(\sqrt{\alpha'}/2) =\mathcal{Z}_g^{S^1/\mathbb{Z}_2}(\sqrt{\alpha'}) \,, \label{c1equiv}
\end{equation}
to derive\footnote{See Appendix \ref{twoloopc1equiv} for the derivation of $T[\zeta_1]$ using this equivalence.}
\begin{equation}
T[\zeta_1] =\left | \frac{\theta [\delta_1] ( \Omega) \, \theta [\delta_1+\zeta_1 ](\Omega)} { \mathcal{Z}_0 (\Omega) \,  \vartheta_3 (\pi_1)^2} \right | \Lambda_{1,1}(R;\pi_1) \,.
\label{eq:twistz1}
\end{equation}
Here $\delta_1 = \twoch{0}{0}{0}{0}$ denotes one of the even characteristics of genus-two theta functions. Notice that the expression of $T[\zeta_1]$ is not unique, since genus-two theta functions satisfy the \emph{Schottky relations}\footnote{See Appendix \ref{app:genus2} for the definition of genus-two theta functions and their Schottky relations.} \cite{fay1973theta}.

\begin{table}
\centering
\begin{tabular}{c|ccccccccccccccc}
    $i$ & 1 & 2 & 3 & 4 & 5 & 6 & 7 & 8 & 9 & 10 & 11 & 12 & 13 & 14 & 15
    \\
    \hline
    $\delta_i$ & 1 & 4 & 9 & 9 & 4 & 2 & 1 & 1 & 5 & 6 & 5 & 6 & 2 & 1 & 1
\end{tabular}
\caption{The association between the twists $\zeta_i$ and the even spin structures $\delta_i$ of the genus-two theta functions.}
\label{tab:spinstructures}
\end{table}

Altogether, the full genus-two partition function for the $T^6/\mathbb{Z}_2^2$ orbifold reads
\begin{equation}
\begin{aligned}
\mathcal{Z}_2 =& \frac{1}{16}\Bigg[ \prod_{v=1,2,3} \Lambda_{2,2}^{(v)} (\Omega)
+ \sum_{v=1,2,3}  \Lambda_{2,2}^{(v)} (\Omega) \sum_{i=1}^{15} \left| \frac{\theta [\delta_i] \, \theta [ \delta_i + \zeta_i]}{\mathcal{Z}_0(\Omega)^2}\, \frac{\eta (\pi_i)^2}{ \vartheta_3 (\pi_i)^2} \right|^4 \, \prod_{u\neq v}\Lambda_{2,2}^{(u)} (\pi_i)
    \\
    &+ \sum_{\substack{i,j \\ i \neq j }} \left| \frac{\theta [ \delta_i] \, \theta [ \delta_i + \zeta_i]}{\mathcal{Z}_0(\Omega)^2} \,\frac{\eta (\pi_i)^2}{ \vartheta_3 (\pi_i)^2} \right|^2 \Lambda_{2,2}^{(1)}(\pi_i) \  \left| \frac{\theta [\delta_j] \, \theta [ \delta_j + \zeta_j]}{\mathcal{Z}_0(\Omega)^2} \, \frac{\eta (\pi_j )^2}{\vartheta_3 (\pi_j)^2} \right|^2  \Lambda_{2,2}^{(3)}(\pi_j) 
    \\
    & \qquad \qquad \times \left| \frac{\theta [ \delta_{i+j}]\, \theta [ \delta_{i+j} + \zeta_i+\zeta_j]}{\mathcal{Z}_0(\Omega)^2} \,\frac{\eta (\pi_{i+j})^2}{\vartheta_3 (\pi_{i+j})^2} \right|^2  \Lambda_{2,2}^{(2)}(\pi_{i+j})\Bigg] \,.
\end{aligned}\label{eq:genus2complete}
\end{equation}
In this expression, $\delta_{i+j}$ and $\pi_{i+j}$ denote the spin structures and modulus of the Prym variety associated to the twist $\zeta_i + \zeta_j$. The association between spin structures $\delta_i$ and twists $\zeta_i$ is given in table \ref{tab:spinstructures}. The Narain lattice $\Lambda_{2,2}^{(v)} (\Omega)$ encodes the contributions of the periodic bosons associated to the $v$-th pair of directions left invariant by the $\mathbb{Z}_2 \times \mathbb{Z}_2$ twists. Again, due to the Schottky relations, the expression \eqref{eq:genus2complete} is not unique.

The partition function \eqref{eq:genus2complete} is modular invariant and, actually, decomposes into six independent modular orbits. To identify them, we note that the modular group $\text{Sp} (4;\mathbb{Z} )$ acts linearly on the twists $\zeta_i$. As a result, $\zeta_0$ is invariant while the $\zeta_{i\neq 0}$ are, in principle, mapped into each others. It can be shown \cite{Aoki:2003sy} that the inner product
\begin{equation}
    \langle \zeta_i, \zeta_j \rangle= e^{4\pi i( \zeta_i^a \cdot \zeta_j^b-\zeta_i^b \cdot \zeta_j^a)}  \,, \label{genus2invprod}
\end{equation}
is invariant under the action of $\text{Sp} (4;\mathbb{Z} )$, and can be naturally used to identify the various modular orbits. In eq. \eqref{genus2invprod},  $\zeta_i^a$ ($\zeta^b_i$) corresponds to the first (second) row of $\zeta_i$, \emph{i.e.} to the periodicities associated to the $a$ ($b$) cycles of $\Sigma_2$ and, for the twists \eqref{eq:twists}, $\langle \zeta_i , \zeta_j \rangle =\pm 1$. This said, we can identify the following orbits.
\begin{itemize}
\item[${\mathcal O}_1$:] This is the orbit associated to the single term $i=0$ and $j=0$ in \eqref{eq:genus2} and has $\langle \zeta_0 , \zeta_0 \rangle = +1$. This contribution is trivially modular invariant since $\zeta_0$ is left invariant under the action of $\text{Sp} (4; \mathbb{Z})$. It is given by
\begin{equation}
\prod_{v=1,2,3} \Lambda^{(v)}_{2,2} (\Omega )\,.
\end{equation}
\item[${\mathcal O}_2$:] This orbit includes the terms $i\neq 0$ and $j=0$ in \eqref{eq:genus2}, with inner product $\langle \zeta_i , \zeta_0\rangle = +1$. Under the action of the modular group, $\zeta_0$ is invariant while the $\zeta_i$ are transformed among themselves. As a result, this orbit has length fifteen, and is associated to the non-trivial $\mathbb{Z}_2$ twist generated by $h$,
\begin{equation}
\Lambda_{2,2}^{(1)} (\Omega) \sum_{i=1}^{15} \left| \frac{\theta [\delta_i] \, \theta [ \delta_i + \zeta_i]}{\mathcal{Z}_0(\Omega)^2}\, \frac{\eta (\pi_i)^2}{ \vartheta_3 (\pi_i)^2} \right|^4 \, \prod_{u\neq 1}\Lambda_{2,2}^{(u)} (\pi_i)\,.
\end{equation}
\item[${\mathcal O}_3$:] This orbit is specular to ${\mathcal O}_2$, and includes the terms $i= 0$ and $j\neq0$ in \eqref{eq:genus2}, with inner product $\langle \zeta_0 , \zeta_j\rangle = +1$. Also this orbit has length fifteen, and is associated to the generator $g$, 
\begin{equation}
\Lambda_{2,2}^{(3)} (\Omega) \sum_{i=1}^{15} \left| \frac{\theta [\delta_i] \, \theta [ \delta_i + \zeta_i]}{\mathcal{Z}_0(\Omega)^2}\, \frac{\eta (\pi_i)^2}{ \vartheta_3 (\pi_i)^2} \right|^4 \, \prod_{u\neq 3}\Lambda_{2,2}^{(u)} (\pi_i)\,.
\end{equation}
\item[${\mathcal O}_4$:] This orbit includes the terms $i=j\neq 0$ in \eqref{eq:genus2}, with inner product $\langle \zeta_i , \zeta_i\rangle = +1$. Since the action of the modular group on the $\zeta_i$'s is linear, the inner product will stay diagonal and thus defines an independent orbit of length fifteen. This is associated to the element $gh=f$, and reads
\begin{equation}
\Lambda_{2,2}^{(2)} (\Omega) \sum_{i=1}^{15} \left| \frac{\theta [\delta_i] \, \theta [ \delta_i + \zeta_i]}{\mathcal{Z}_0(\Omega)^2}\, \frac{\eta (\pi_i)^2}{ \vartheta_3 (\pi_i)^2} \right|^4 \, \prod_{u\neq 2}\Lambda_{2,2}^{(u)} (\pi_i)\,.
\end{equation}
\item[${\mathcal O}_5$:] This orbit includes the terms with $i\neq j$ and positive inner product, $\langle \zeta_i , \zeta_j \rangle =+1$. To determine the structure of this orbit we start from the inner product $\langle \zeta_1 , \zeta_j\rangle$ and identify the $\zeta_j$ such that $\langle \zeta_1 , \zeta_j\rangle =+1$. There are 6 such $\zeta$'s with $j=10, \ldots ,15$. Modular transformations map  $\langle \zeta_1 , \zeta_j\rangle \to \langle \zeta_k , \zeta_{\ell}\rangle $ generating $15\times 6 = 90$ terms, all with positive inner product. As a result, this orbit has length ninety and is given by
\begin{equation}
\begin{aligned}
    &\sum_{\substack{i \neq j \\  \langle \zeta_i, \zeta_j \rangle =+1 }} \left| \frac{\theta [ \delta_i] \, \theta [ \delta_i + \zeta_i]}{\mathcal{Z}_0(\Omega)^2} \,\frac{\eta (\pi_i)^2}{ \vartheta_3 (\pi_i)^2} \right|^2 \Lambda_{2,2}^{(1)}(\pi_i) \  \left| \frac{\theta [\delta_j] \, \theta [ \delta_j + \zeta_j]}{\mathcal{Z}_0(\Omega)^2} \, \frac{\eta (\pi_j )^2}{\vartheta_3 (\pi_j)^2} \right|^2  \Lambda_{2,2}^{(3)}(\pi_j) 
    \\
    & \qquad \qquad \times \left| \frac{\theta [ \delta_{i+j}]\, \theta [ \delta_{i+j} + \zeta_i+\zeta_j]}{\mathcal{Z}_0(\Omega)^2} \,\frac{\eta (\pi_{i+j})^2}{\vartheta_3 (\pi_{i+j})^2} \right|^2  \Lambda_{2,2}^{(2)}(\pi_{i+j})  \, .
\end{aligned}\label{eq:O5orbit}
\end{equation}
\item[${\mathcal O}_6$:] Finally, we are left with the $\zeta$'s such that $\langle \zeta_i , \zeta_j \rangle = - 1$. As before, we start from those twist vectors which have $\langle \zeta_1 , \zeta_j \rangle = - 1$. These have $j=2,\ldots ,9$. Modular transformations generate all the others. This orbit has then length $15 \times 8 = 120$ and reads
\begin{equation}
\begin{aligned}
& \sum_{\substack{i \neq j \\ \langle \zeta_i, \zeta_j \rangle =-1 }} \left| \frac{\theta [ \delta_i] \, \theta [ \delta_i + \zeta_i]}{\mathcal{Z}_0(\Omega)^2} \,\frac{\eta (\pi_i)^2}{ \vartheta_3 (\pi_i)^2} \right|^2 \Lambda_{2,2}^{(1)}(\pi_i) \  \left| \frac{\theta [\delta_j] \, \theta [ \delta_j + \zeta_j]}{\mathcal{Z}_0(\Omega)^2} \, \frac{\eta (\pi_j )^2}{\vartheta_3 (\pi_j)^2} \right|^2  \Lambda_{2,2}^{(3)}(\pi_j) 
    \\
    & \qquad \qquad \times \left| \frac{\theta [ \delta_{i+j}]\, \theta [ \delta_{i+j} + \zeta_i+\zeta_j]}{\mathcal{Z}_0(\Omega)^2} \,\frac{\eta (\pi_{i+j})^2}{\vartheta_3 (\pi_{i+j})^2} \right|^2  \Lambda_{2,2}^{(2)}(\pi_{i+j})\, .
    \end{aligned}
\end{equation} 
\end{itemize}

As in the one-loop case, these independent orbits can support discrete torsion. We can fix the sign of ${\mathcal O}_1$ to be positive. The sign ambiguity of the orbits ${\mathcal O}_{2,3,4}$ is also fixed, since each orbit defines a proper $\mathbb{Z}_2$ projection of ${\mathcal O}_1$. So, we are left, in principle, with two arbitrary signs associated to ${\mathcal O}_5$ and ${\mathcal O}_6$. As we shall see, this is not completely true since they are fixed by the discrete torsion one can introduce at one-loop.
 This is a consequence of a crucial property of genus-two amplitudes: the period matrix
\begin{equation}
\Omega= \begin{pmatrix} \tau_1 & t \\  t & \tau_2\end{pmatrix} \,,
\label{eq:Omegaparametrisation}
\end{equation}
describes, for small $t$,  two tori $\Sigma_1$ connected by a tube of length $-\log t$ and, in the $t\to 0$ limit, factorises as $\Sigma_2 (\Omega) \to \Sigma_1 (\tau_1 ) \times \Sigma_1 (\tau_2)$. Then, using the factorisation properties of theta functions reported in Appendix \ref{app:genus2}, it is straightforward to get
\begin{equation}
\begin{aligned}
 {\mathcal O}_1  &\to  \mathcal{Z} \onech{1}{1}(\tau_1)  \ \mathcal{Z} \onech{1}{1}(\tau_2) \,,
 \\
 \mathcal{O}_2 &\to  \mathcal{Z} \onech{1}{1}(\tau_1) {\mathcal Z}_h (\tau_2 ) + {\mathcal Z}_h (\tau_1 ) \mathcal{Z} \onech{1}{1}(\tau_2) + {\mathcal Z}_h (\tau_1 ) {\mathcal Z}_h (\tau_2 )\,,
\\
\mathcal{O}_3 &\to  \mathcal{Z} \onech{1}{1}(\tau_1) {\mathcal Z}_g (\tau_2 ) + {\mathcal Z}_g (\tau_1 ) \mathcal{Z} \onech{1}{1}(\tau_2) + {\mathcal Z}_g (\tau_1 ) {\mathcal Z}_g (\tau_2 )\,,
\\
\mathcal{O}_4 &\to  \mathcal{Z} \onech{1}{1}(\tau_1) {\mathcal Z}_f (\tau_2 ) + {\mathcal Z}_f (\tau_1 ) \mathcal{Z} \onech{1}{1}(\tau_2) + {\mathcal Z}_f (\tau_1 ) {\mathcal Z}_f (\tau_2 ) \,,
\\
\mathcal{O}_5  &\to \sum_{\substack{\alpha ,\beta = g,h,f \\ \alpha \neq \beta}} {\mathcal Z}_\alpha (\tau_1 ) {\mathcal Z}_\beta (\tau_2 ) + {\mathcal Z}_{g,h} (\tau_1 ) {\mathcal Z}_{g,h} (\tau_2 ) \,,
\\
{\mathcal O}_6 &\to {\mathcal Z}_{g,h} (\tau_1 ) \left( \mathcal{Z} \onech{1}{1}(\tau_2) +\sum_{\alpha = g,h,f} {\mathcal Z}_\alpha (\tau_2 ) \right) + \left( \mathcal{Z} \onech{1}{1}(\tau_1) +\sum_{\alpha = g,h,f} {\mathcal Z}_\alpha (\tau_1 ) \right) {\mathcal Z}_{g,h} (\tau_2 )  \,.
\end{aligned}
\end{equation}
In these expressions ${\mathcal Z} \onech{1}{1}$ refers to the one-loop partition function associated to the untwisted unprojected sector, ${\mathcal Z}_\alpha = {\mathcal Z}\onech{1}{\alpha} +  {\mathcal Z}\onech{\alpha}{1} + {\mathcal Z}\onech{\alpha}{\alpha} $ is the (one-loop) modular orbit associated to the untwisted amplitude with the $\alpha$ generator inserted in the trace, while
\begin{equation}
{\mathcal Z}_{g,h} =  {\mathcal Z}\onech{g}{h} + {\mathcal Z}\onech{g}{f} + {\mathcal Z}\onech{h}{g}+ {\mathcal Z}\onech{h}{f} + {\mathcal Z}\onech{f}{g} + {\mathcal Z}\onech{f}{h} \,,
\end{equation}
is the independent (one-loop) modular orbit completing the orbifold projection in the twisted sectors.

The generalised discrete torsion of \cite{Gaberdiel:2004vx} is compatible with the modular invariance of the ${\mathcal O}_6$ orbit. In fact, at one-loop, a point-wise discrete torsion $\epsilon_r$ affects the structure of twisted Hilbert spaces as in eq. \eqref{eq:gdt}, and  implies
\begin{equation}
{\mathcal Z}_{g,h} = \sum_{r=1}^{16} \, \epsilon_r \sum_{\substack{\alpha \neq 1 \\ \beta\neq (1,\alpha)}} \hat {\mathcal Z} \onech{\alpha}{\beta} \,. \label{eq:gdt4}
\end{equation}
As a result, the genus-two point-wise discrete torsion $\sum_{r=1}^{16} \epsilon_r$ for the ${\mathcal O}_6$ orbit, would reproduce the correct ${\mathcal Z}_{g,h}$ in the factorisation limit.

This is not the case for the ${\mathcal O}_5$ orbit, since different pairs $\zeta_i$ and $\zeta_j$ yield different terms in the factorisation limit. For instance, $\zeta_1$ and $\zeta_{10}$ contribute to terms of the type ${\mathcal Z}_\alpha \, {\mathcal Z}_\beta$, while $\zeta_2$ and $\zeta_{10}$ contribute to ${\mathcal Z}_{g,h}^2$. To reproduce the one-loop generalised discrete torsion \eqref{eq:gdt4}, one should introduce in eq. \eqref{eq:O5orbit} different coefficients for the various $\zeta$'s. But, this is clearly incompatible with  ${\mathcal O}_5$ being an independent, irreducible, orbit. Therefore,  two-loop modular invariance requires 
\begin{equation}
\left(  \frac{1}{16}\, \sum_{r=1}^{16} \epsilon_r \right)^2 =1\,,
\end{equation}
thus implying that only the standard discrete torsion of \cite{Vafa:1986wx} be allowed.

To reiterate, let us assume that one can introduce the two-loop generalised discrete torsion $\epsilon(\alpha_1,\beta_1,r_1;\alpha_2,\beta_2,r_2)$ on the fixed points $r_i$, with $\alpha_i$ and $\beta_i$ being the orbifold elements acting on $\Sigma_2$. Its transformation 
\begin{equation}
     \epsilon(\alpha_1,\beta_1,r_1;\alpha_2,\beta_2,r_2)=\epsilon(\alpha_1 \beta_2 \beta_1^{-1},\beta_1,r_1';\alpha_2 \beta_1 \beta_2^{-1},\beta_2,r_2') \, ,
\end{equation}
under the action of the Dehn twist
\begin{equation}
\Sigma ' =\begin{pmatrix} 1 & 0 & 0 & 0 \\ 0 & 1 & 0 & 0 \\ -1 & 1 & 1 & 0 \\ 1 & -1 & 0 & 1\end{pmatrix}\,, 
\label{eq:Dehn}
\end{equation}
together with the factorisation property
\begin{equation}
    \epsilon(\alpha_1,\beta_1,r_1;\alpha_2,\beta_2,r_2)=\epsilon(\alpha_1,\beta_1,r_1) \, \epsilon(\alpha_2,\beta_2,r_2) \, ,
    \label{eq:factor}
\end{equation}
implies, in particular,
\begin{equation}
\epsilon(g,g,r_1;f,f,r_2)=\epsilon(g,g,r_1) \epsilon(f,f,r_2)=1=\epsilon(f,g,r_1';g,f,r_2')= \epsilon(f,g,r_1')\epsilon(g,f,r_2') \, .
\end{equation}
Under the reasonable assumption that the generic fixed points $r_1$ and $r_2$ are not all mapped to a single one under the action of the Dehn twist, but are, at most, permuted among themselves, the only solution is
\begin{equation}
    \epsilon(f,g,r_1')=\epsilon(g,f,r_2') = \epsilon (f,g, r_2') \qquad \text{for\ any}\quad r_1^\prime \,, \ r_2^\prime\,,
\end{equation}
and thus only the standard discrete torsion is allowed.

To conclude, the operator interpretation of the one-loop vacuum energy and the modular invariance of the genus-two partition function forbid any local choice \cite{Gaberdiel:2004vx} of discrete torsion.

\subsection{Comments on the $T^7/\mathbb{Z}_2^3$ orbifold}

The discussion of the $T^7/\mathbb{Z}_2^3$ orbifold follows, in principle, a similar pattern, although the presence of the shift in the generators \eqref{G2gen} induces some complications. The two loop partition function can be schematically written as
\begin{equation}
{\mathcal Z}_2 = \frac{1}{2^{6}} \sum_{i,j,k=0}^{15} T[\zeta_i+\zeta_j +\zeta_k | \sigma_k] \, T [\zeta_i+ \zeta_j |\sigma_j] \, T [\zeta_i+\zeta_k]\, T [\zeta_i]\, T [\zeta_j+\zeta_k] \, T [\zeta_j] \, T [\zeta_k] \,, \label{eq:2lg2}
\end{equation}
where, as before, each $T$ refers to a given coordinate twisted by a $\zeta$ in eq. \eqref{eq:twists} and, possibly, shifted by $\sigma$. The amplitude $T [\zeta | \sigma]$, with a non-trivial $\zeta$ involves the action of the shift $\sigma$ on the Prym variety which, to the best of our knowledge, is not known. Despite this difficulty, the (partial) orbit decomposition of ${\mathcal Z}_2$ and its (formal) factorisation suffice to draw our conclusions. In fact, among the 4096 terms in eq. \eqref{eq:2lg2} we can extract
\begin{equation}
\begin{aligned}
{\mathcal Z}_2 &= \frac{1}{2^{6}} \Bigg[ T[\zeta_0]^7 + \sum_{i=1}^{15} T[\zeta_i]^4 \, T [\zeta_0]^3 + \ldots +
\\
& + \sum_{\substack{i=j\neq k \\ \langle \zeta_j , \zeta_k \rangle =1}} T[\zeta_k | \sigma_k] \, T [\zeta_0 |\sigma_j] \, T [\zeta_j+\zeta_k]\, T [\zeta_j]\, T [\zeta_j+\zeta_k] \, T [\zeta_j] \, T [\zeta_k] +\ldots \Bigg] \,.
\end{aligned}
\end{equation}
The first term corresponds to the trivial orbit and factorises as ${\mathcal Z}\onech{1}{1} (\tau_1 )\, {\mathcal Z}\onech{1}{1} (\tau_2 )$. The second term is associated to the $a$ generator in eq. \eqref{G2gen} and, along the lines of the previous Section, factorises as ${\mathcal Z}\onech{1}{1} (\tau_1 )\, {\mathcal Z}_a (\tau_2 ) + {\mathcal Z}_a (\tau_1 )\, {\mathcal Z}\onech{1}{1} (\tau_2 ) + {\mathcal Z}_a (\tau_1 )\, {\mathcal Z}_a (\tau_2 )$. The third term is the most interesting one since it is associated to the non-trivial orbit, with positive inner product, of a $\mathbb{Z}_2 \times \mathbb{Z}_2$ orbifold generated by $c$ and $ab$. As represented in figure \ref{G2}, this is the only $\mathbb{Z}_2^2 \subset \mathbb{Z}_2^3$ which can support discrete torsion. This orbit then factorises as
\begin{equation}
\sum_{\substack{\alpha , \beta = c, ab, abc \\ \alpha \neq \beta}} {\mathcal Z}_\alpha (\tau_1 ) \, {\mathcal Z}_\beta (\tau_2 ) + {\mathcal Z}_{c,ab} (\tau_1 )  {\mathcal Z}_{c,ab} (\tau_2 ) \,,
\end{equation}
with ${\mathcal Z}_{c,ab}$ the one-loop modular orbit associated to the grey boxes in figure \ref{G2}. This factorisation has the same structure as the decomposition of  ${\mathcal O}_5$  of the previous Section, and therefore leads to the same conclusion, \emph{i.e.} two-loop modular invariance is only compatible with the standard discrete torsion of \cite{Vafa:1986wx}. In fact, introducing a two-loop generalised discrete torsion $\epsilon (\alpha_1 , \beta_1 , r_1 ; \alpha_2 , \beta_2 , r_2 )$, its transformation under the Dehn twist \eqref{eq:Dehn} and the factorisation \eqref{eq:factor} imply
\begin{equation}
    \epsilon(a b, a b ,r_1; c,c,r_2)=\epsilon( ab, ab,r_1) \epsilon(c,c,r_2)=1=\epsilon(c,ab ,r_1';ab,c,r_2')= \epsilon(c,ab,r_1')\epsilon(ab,c,r_2') \, ,
\end{equation}
which leaves no room for a generalised discrete torsion.

\bigskip

\acknowledgments

We are grateful to Ivano Basile and Luca Martucci for enlightening discussions. G.L. is grateful to Salvatore Raucci for insightful discussions and collaboration on related projects. C.A. and G.L. would like to thank the Theory Department at CERN and  D.P. would like to thank the Physics Department in Torino for hospitality during various stages of this work. The work of C.A. is partly supported by the Higher Education and Science Committee of RA (Research Project N 24RL-1C036). The work of G.L. is funded by the European Union - NextGenerationEU/PNRR mission 4.1; CUP: C93C24004950006.
\\

\noindent
In no way the artificial intelligence has played any role in the analysis performed in this paper.

\appendix

\section{Theta functions on Riemann surfaces}\label{app:genus2}

On a genus-$g$ Riemann surface $\Sigma_g$ with period matrix $\Omega$ in the Siegel upper-half space $\mathbb{H}_g$, one can introduce the theta functions
\begin{equation}\label{eq:theta}
    \theta[\rho ](z|\Omega)= \sum_{m \in \mathbb{Z}^g} e^{ i \pi \, (m+\rho^a)^t \Omega(m+\rho^a)}\, e^{2 \pi i  \, (m+\rho^a)^t (z+\rho^b)} \, ,
\end{equation}
with characteristics
\begin{equation}
\rho = \onech{\rho^a}{\rho^b} = \Big[{\textstyle{{\rho^{a_1} \atop \rho^{b_1}} \ldots {\rho^{a_g} \atop \rho^{b_g}}}}\Big]\,,
\end{equation}
along the $a$ and $b$ cycles of $\Sigma_g$. For $\ell , k \in \mathbb{Z}^g$, the theta functions are clearly invariant under the shift  $\rho^a \to \rho^a + \ell$ of the upper characteristic, while they pick-up the phase $e^{2 \pi i \, \rho^a \cdot k}$  under the shift $\rho^b \to \rho^b + k$ of the lower characteristics. For vanishing or half-integer characteristic $\rho$ the theta functions have a definite parity under the reflection $z\to - z$
\begin{equation}
\theta [\rho ] (- z |\Omega ) = (-1)^{4\, \rho^a \cdot \rho^b}\,  \theta [\rho ] (z | \Omega) \,.
\end{equation}
We call the characteristic \emph{even} (\emph{odd}) if  $4 \, \rho^a \cdot \rho^b =0 \, (\text{mod}\, 2)$ ($4 \, \rho^a \cdot \rho^b =1 \, (\text{mod}\, 2)$). On $\Sigma_g$ there are exactly $2^{g-1} (2^g \pm 1)$ theta functions with even/odd characteristics.

Other useful properties of the theta functions are
\begin{equation}
\theta \onech{-\rho^a}{\rho^b} (z|\Omega ) = \theta \onech{\rho^a}{-\rho^b} (-z|\Omega ) \,,
\end{equation}
\begin{equation}
\theta \onech{\rho^a}{\rho^b} (z| 4\, \Omega ) = \frac{1}{2^g}\sum_\gamma e^{-4 \pi i \rho^a \cdot \gamma}\, \theta \onech{2\rho^a}{\gamma+\rho^b/2} (z/2|\Omega) \,,
\end{equation}
together with the duplication formula
\begin{equation}
\theta \onech{\alpha}{\gamma} (2 z_1 | 2 \Omega ) \, \theta \onech{\beta}{\gamma} (2 z_2 |2\Omega ) = \frac{1}{2^g} \sum_\epsilon (-1)^{4 \alpha \cdot \epsilon } \, \theta\onech{\alpha+\beta}{\gamma+\epsilon} (z_1 + z_2 |\Omega)\, \theta \onech{\alpha+\beta}{\epsilon} (z_1 - z_2 |\Omega) \,, \label{eq:duplication}
\end{equation}
valid for half-integer characteristics. Most important for our purposes are the Schottky relations
\begin{equation}
\frac{\theta \twoch{\rho}{0}{0}{\nicefrac{1}{2}} (0|2\Omega )^2}{\theta \onech{\rho}{0} (0|2\pi )^2}= \frac{\theta\twoch{\alpha}{0}{\beta}{0} (0|\Omega)\, \theta\twoch{\alpha}{0}{\beta}{\nicefrac{1}{2}} (0|\Omega ) }{\theta^2 \onech{\alpha}{\beta} (0|\pi )}\,, \label{eq:Schottky}
\end{equation}
valid for half-integer characteristics $\alpha, \beta, \rho$, and $\pi$ the period of the Prym variety \cite{fay1973theta}.

The period matrices $\Omega$ and $\Omega '$ describing equivalent $\Sigma_g$'s are connected by the projective transformation
\begin{equation}
\Omega ' = (A \, \Omega + B )\, (C \, \Omega + D)^{-1}\,, \qquad \text{with} \quad M=\begin{pmatrix} A & B \\ C & D \end{pmatrix} \in \text{Sp} (2 g ; \mathbb{Z})\,.
\end{equation}
Under such $M$ the theta functions, evaluated at $z=0$, transform as
\begin{equation}
        \theta[\rho ' ](\Omega ' )= \varepsilon (M)\, \det(C\Omega+D)^{\frac12} \, \theta[\rho ](\Omega)  \, ,
\end{equation}
where the multiplicative system $\varepsilon (M)^8=1$, and
\begin{equation}
\begin{pmatrix} \rho^{\prime\, a} \\ \rho^{\prime \, b} \end{pmatrix}= 
\begin{pmatrix}   D  & -C  \\   -B & A  \end{pmatrix} \,
\begin{pmatrix} \rho^a \\ \rho^b \end{pmatrix} 
+\frac{1}{2}\,  \, \begin{pmatrix} \text{diag} (C D^t)  \\ \text{diag} (A B^t)  \end{pmatrix}  \,.
\end{equation}
Therefore, even and odd theta functions never mix under the action of the modular group $\text{Sp} (2 g ; \mathbb{Z})$. 
 
In this paper, we are particularly interested in theta functions on $\Sigma_1$ and on $\Sigma_2$. At genus one one has the familiar Jacobi theta functions
\begin{equation}
\begin{aligned}
\theta\onech{0}{0} (z|\tau ) &\equiv \vartheta_3 (z|\tau)\,, \\ \theta \onech{0}{\nicefrac{1}{2}} (z|\tau ) &\equiv \vartheta_4 (z|\tau)\,, 
\end{aligned}\qquad  
\begin{aligned}
\theta\onech{\nicefrac{1}{2}}{0} (z|\tau ) &\equiv \vartheta_2 (z|\tau ) \,, \\ \theta\onech{\nicefrac{1}{2}}{\nicefrac{1}{2}} (z|\tau ) &\equiv -\vartheta_1 (z|\tau)\,,
\end{aligned}
\end{equation}
with $\vartheta_2$, $\vartheta_3$ and $\vartheta_4$ even, while $\vartheta_1$ is odd.
They obey the triple product relation 
\begin{equation}
\vartheta_2 (0 |\tau) \, \vartheta_3 (0|\tau ) \, \vartheta_4 (0|\tau) = 2 \eta^3 (\tau)\,, \qquad \text{with}\quad \eta (\tau) = e^{i \pi \tau/12} \, \prod_{n=1}^\infty (1 - e^{2 \pi i n \tau})\,,
\end{equation}
instrumental for fermionisation, and the Jacobi \emph{{\ae}quatio identica satis abstrusa}
\begin{equation}
\vartheta_3^4 (0|\tau ) - \vartheta_4^4 (0|\tau ) - \vartheta_2 ^4 (0|\tau ) \equiv 0\,,
\end{equation}
strictly linked to the presence of space-time supersymmetry. The generators of the modular group $\text{Sp} (2;\mathbb{Z}) \simeq \text{PSL} (2;\mathbb{Z})$ are
\begin{equation}
T = \begin{pmatrix} 1 & 1 \\ 0 & 1 \end{pmatrix}\qquad \text{and}\qquad
S = \begin{pmatrix} 0 & 1 \\ -1 & 0 \end{pmatrix}\,.
\end{equation}

At genus two, there are ten theta functions with even characteristics
\begin{equation}\label{eq:evengenus2}
\begin{aligned} 
\delta_1&=\twoch{0}{0}{0}{0}\,, 
\\
\delta_6 &= \twoch{\nicefrac{1}{2}}{\nicefrac{1}{2}}{0}{0}\,,
\end{aligned}    
\quad 
\begin{aligned}    
\delta_2 &= \twoch{0}{0}{\nicefrac{1}{2}}{\nicefrac{1}{2}} \,,
\\
\delta_7 &= \twoch{\nicefrac{1}{2}}{\nicefrac{1}{2}}{\nicefrac{1}{2}}{\nicefrac{1}{2}}\,,
\end{aligned}
\quad
\begin{aligned}
\delta_3&=\twoch{0}{0}{\nicefrac{1}{2}}{0}\,,
\\
\delta_8 &=\twoch{\nicefrac{1}{2}}{0}{0}{0}\,,
\end{aligned}
\quad
\begin{aligned}
\delta_4 &=\twoch{0}{0}{0}{\nicefrac{1}{2}}\,,
\\
\delta_9 &= \twoch{0}{\nicefrac{1}{2}}{0}{0}\,,
\end{aligned}
\quad
\begin{aligned}
\delta_5 &=\twoch{0}{\nicefrac{1}{2}}{\nicefrac{1}{2}}{0}\,,
\\
\delta_{10} &= \twoch{\nicefrac{1}{2}}{0}{0}{\nicefrac{1}{2}}\,,
\end{aligned}
\end{equation}
and six with odd characteristics
\begin{equation}
\begin{aligned}\label{eq:oddspinstructures}
\nu_1 &=\twoch{\nicefrac{1}{2}}{0}{\nicefrac{1}{2}}{0}\,,
\\
\nu_4 &= \twoch{0}{\nicefrac{1}{2}}{0}{\nicefrac{1}{2}}\,,
\end{aligned}
\quad
\begin{aligned}
\nu_2 &= \twoch{\nicefrac{1}{2}}{0}{\nicefrac{1}{2}}{\nicefrac{1}{2}}\,,
\\
\nu_5 &= \twoch{0}{\nicefrac{1}{2}}{\nicefrac{1}{2}}{\nicefrac{1}{2}}\,,
\end{aligned}
\quad
\begin{aligned}
\nu_3 &= \twoch{\nicefrac{1}{2}}{\nicefrac{1}{2}}{\nicefrac{1}{2}}{0}\,,
\\
\nu_6 &= \twoch{\nicefrac{1}{2}}{\nicefrac{1}{2}}{0}{\nicefrac{1}{2}} \,.
\end{aligned}
\end{equation}

The generators of the modular group $\text{Sp} (4;\mathbb{Z})$ are
\begin{equation}\label{eq:S1T1}
T_1=\begin{pmatrix}
        1 & 0 & 1 & 0
        \\
        0 & 1 & 0 & 0
        \\
        0 & 0 & 1 &0
        \\
        0 & 0 &0 & 1
    \end{pmatrix} \, , \quad 
    S_1=\begin{pmatrix}
        0 & 0 & 1 & 0
        \\
        0 & 1 & 0 & 0
        \\
        -1 & 0 & 0 &0
        \\
        0 & 0 &0 & 1
    \end{pmatrix} \, , \quad 
 \quad T_2=\begin{pmatrix}
        1 & 0 &0 & 0
        \\
        0 & 1 & 0 & 1 
        \\
        0 & 0 & 1 & 0 
        \\
        0 & 0 & 0 & 1 
    \end{pmatrix} \, , \quad
    S_2=\begin{pmatrix}
        1 & 0 &0 & 0
        \\
        0 & 0 & 0 & 1 
        \\
        0 & 0 & 1 & 0 
        \\
        0 & -1 & 0 & 0 
    \end{pmatrix} \, ,
\end{equation}
and 
\begin{equation}\label{eq:Sigma}
    \Sigma'= \begin{pmatrix}
        1 & 0 & 0 & 0
        \\
        0 & 1 & 0 & 0
        \\
        -1 & 1 & 1 & 0
        \\
        1 & -1 &0 & 1
    \end{pmatrix} \, .
\end{equation}
Each pair $T_i , S_i$ generates a $\text{PSL} (2;\mathbb{Z})$ subgroup which rotates \emph{diagonally} the $a_i$ and $b_i$ cycles, \emph{i.e.} without mixing $a_1,b_1$ with $a_2 , b_2$. For the special choice of period matrix
\begin{equation}
\Omega = \begin{pmatrix} \tau_1 & 0 \\ 0 &\tau_2 \end{pmatrix}\,,
\end{equation}
they act projectively on $\tau_i$. The Dehn twist $\Sigma '$, instead, connects the two pairs of cycles as
\begin{equation}
\twoch{a_1}{a_2}{b_1}{b_2} \to \twoch{a_1}{a_2}{b_1 - a_1 + a_2}{b_2 +a_1 - a_2}\,.
\end{equation}

Finally, the factorisation limit of the period matrix
\begin{equation}
\Omega = \begin{pmatrix} \tau_1 & t \\ t & \tau_1 \end{pmatrix}\,,
\end{equation}
as $t\to 0$ implies the following decomposition of the even genus-two theta functions at zero argument in terms of genus-one $\theta$'s
\begin{equation}
  \begin{aligned} \label{eq:thetafactorisation}
    \theta [  \delta_{i\neq 7}](0|\Omega) &= \theta[\mu_{i,1} ](0|\tau_1) \,  \theta[\mu_{i,2}](0|\tau_2) + O(t^2) \, ,
    \\
    \theta [ \delta_7 ](0|\Omega) &= - \ 2\pi i \, t \,  \eta(\tau_1)^3 \eta(\tau_2)^3 \, + O(t^2) \, , 
    \end{aligned}
\end{equation}
where the $\mu$'s are genus-one even spin structures in $\delta_{i\neq 7} = [\mu_{i,1}\ \mu_{i,2}]$. For instance, 
\begin{equation}
\delta_5 = \twoch{0}{\nicefrac{1}{2}}{\nicefrac{1}{2}}{0} \qquad \Rightarrow \qquad \mu_{5,1} = \onech{0}{\nicefrac{1}{2}}\quad \text{and}\quad  \mu_{5,2} = \onech{\nicefrac{1}{2}}{0}\,,
\end{equation}
and
\begin{equation}
\theta [  \delta_{5}](0|\Omega) = \vartheta_4 (0|\tau_1 ) \, \vartheta_2 (0|\tau_2) + O(t^2)\,.
\end{equation}

\section{Details on one-loop partition functions}\label{AppA}

\subsection{Useful formulae for the $\text{U} (1)_4$ characters}

We summarise here some properties of the $\text{U} (1)_4$ characters used in the main body of the paper. These  correspond to a compact boson at radius $R=\sqrt{4\alpha '}$. There are eight such characters given by
\begin{equation}
\xi_a (q) = \frac{1}{\eta (q)} \sum_{n\in\mathbb{Z}} q^{4 (n +a/8)^2}\,, \qquad a=-3\ldots 4\,, \label{R4ch}
\end{equation}
where we have introduced the \emph{nome} $q=e^{2\pi i\tau}$.
They have conformal weight $h_a = a^2/16$ and, clearly, the identity is given by $\xi_0$. The other real character is $\xi_4$, while $\xi_{\pm n}$ form conjugate pairs. The $T$ and $S$ modular transformations of the $\xi$'s can be straightforwardly obtained by the definition \eqref{R4ch} and are represented by the matrices
\begin{equation}
T_{ab} = e^{2i \pi (  \frac{a^2}{16} - \frac{1}{24})}\delta_{ab}\,, \qquad
S_{ab} = \frac{e^{2 \pi i a b /8}}{2\sqrt{2}}\,.
\end{equation}
Notice that
\begin{equation}
\begin{aligned}
\xi_0 (q)+ \xi_4 (q) &= \frac{1}{\eta(q)} \sum_{n\in\mathbb{Z}} \left(q^{4 n^2} + q^{4 (n+1/2)^2} \right) = \frac{1}{\eta(q)} \sum_{n\in\mathbb{Z}} q^{n^2} \equiv \chi_0 (q)\,,
\\
\xi_2 (q)+ \xi_{-2} (q) &= \frac{1}{\eta(q)} \sum_{n\in\mathbb{Z}} \left(q^{4 (n+1/4)^2} + q^{4 (n-1/4)^2} \right) = \frac{1}{\eta(q)} \sum_{n\in\mathbb{Z}} q^{(n+1/2)^2} \equiv \chi_{\nicefrac{1}{2}} (q)\,,
\end{aligned}
\end{equation}
where $\chi_0$ and $\chi_{\nicefrac{1}{2}}$ are the characters of SU(2) level one. Finally, 
\begin{equation}
\frac{1}{\eta\, \bar\eta}\sum_{m,n} (-1)^m \, q^{\frac{\alpha '}{4} \left( \frac{m}{R} +\frac{nR}{\alpha '}\right)^2}\, \bar q^{\frac{\alpha '}{4} \left( \frac{m}{R} -\frac{nR}{\alpha '}\right)^2} = |\xi_0 - \xi_4|^2 - |\xi_2 - \xi_{-2}|^2\,,
\end{equation}
as explicitly shown in \cite{Angelantonj:2023egh}.

\subsection{The most general point-wise discrete torsion for the $T^6/\mathbb{Z}_2^2$ orbifold}

In the $T^6/\mathbb{Z}_2^2$ orbifold, the amplitude 
\begin{equation}
{\mathcal Z} \onech{g}{1} = \prod_{i\in \{1,2,3,4\}} \left[ |\xi^i_1 + \xi^i_{-3}|^2 + |\xi^i_{-1} + \xi^i_{3}|^2 \right]\, \prod_{i \in \{5,6\}} \left[ |\xi^i_0 + \xi^i_4|^2 + |\xi^i_2 + \xi^i_{-2}|^2 \right]\,,
\end{equation}
encodes the structure of the ground state and its excitations on each fixed point. Indeed, by expanding the product over the twisted coordinates, and barring the contribution of the two untwisted bosons with $i =5,6$, one gets the sixteen monomials 
\begin{equation}
\begin{aligned}
&|(\xi_1^1 + \xi_{-3}^1)(\xi_1^2 + \xi_{-3}^2)(\xi_1^3 + \xi_{-3}^3)(\xi_1^4 + \xi_{-3}^4)|^2 + |(\xi_1^1 + \xi_{-3}^1)(\xi_1^2 + \xi_{-3}^2)(\xi_1^3 + \xi_{-3}^3)(\xi_{-1}^4 + \xi_{3}^4)|^2
\\
+&|(\xi_1^1 + \xi_{-3}^1)(\xi_1^2 + \xi_{-3}^2)(\xi_{-1}^3 + \xi_{3}^3)(\xi_1^4 + \xi_{-3}^4)|^2 + |(\xi_1^1 + \xi_{-3}^1)(\xi_{-1}^2 + \xi_{3}^2)(\xi_1^3 + \xi_{-3}^3)(\xi_1^4 + \xi_{-3}^4)|^2
\\
+&|(\xi_{-1}^1 + \xi_{3}^1)(\xi_1^2 + \xi_{-3}^2)(\xi_1^3 + \xi_{-3}^3)(\xi_1^4 + \xi_{-3}^4)|^2 + |(\xi_1^1 + \xi_{-3}^1)(\xi_1^2 + \xi_{-3}^2)(\xi_{-1}^3 + \xi_{3}^3)(\xi_{-1}^4 + \xi_{3}^4)|^2
\\
+& |(\xi_1^1 + \xi_{-3}^1)(\xi_{-1}^2 + \xi_{3}^2)(\xi_1^3 + \xi_{-3}^3)(\xi_{-1}^4 + \xi_{3}^4)|^2 + |(\xi_{-1}^1 + \xi_{3}^1)(\xi_1^2 + \xi_{-3}^2)(\xi_1^3 + \xi_{-3}^3)(\xi_{-1}^4 + \xi_{3}^4)|^2
\\
+& |(\xi_1^1 + \xi_{-3}^1)(\xi_{-1}^2 + \xi_{3}^2)(\xi_{-1}^3 + \xi_{3}^3)(\xi_1^4 + \xi_{-3}^4)|^2 + |(\xi_{-1}^1 + \xi_{3}^1)(\xi_1^2 + \xi_{-3}^2)(\xi_{-1}^3 + \xi_{3}^3)(\xi_1^4 + \xi_{-3}^4)|^2
\\
+& |(\xi_{-1}^1 + \xi_{3}^1)(\xi_{-1}^2 + \xi_{3}^2)(\xi_1^3 + \xi_{-3}^3)(\xi_1^4 + \xi_{-3}^4)|^2 + |(\xi_{-1}^1 + \xi_{3}^1)(\xi_{-1}^2 + \xi_{3}^2)(\xi_{-1}^3 + \xi_{3}^3)(\xi_1^4 + \xi_{-3}^4)|^2
\\
+&|(\xi_{-1}^1 + \xi_{3}^1)(\xi_{-1}^2 + \xi_{3}^2)(\xi_{1}^3 + \xi_{-3}^3)(\xi_{-1}^4 + \xi_{3}^4)|^2+|(\xi_{-1}^1 + \xi_{3}^1)(\xi_{1}^2 + \xi_{-3}^2)(\xi_{-1}^3 + \xi_{3}^3)(\xi_{-1}^4 + \xi_{3}^4)|^2
\\
+& |(\xi_{1}^1 + \xi_{-3}^1)(\xi_{-1}^2 + \xi_{3}^2)(\xi_{-1}^3 + \xi_{3}^3)(\xi_{-1}^4 + \xi_{3}^4)|^2 + |(\xi_{-1}^1 + \xi_{3}^1)(\xi_{-1}^2 + \xi_{3}^2)(\xi_{-1}^3 + \xi_{3}^3)(\xi_{-1}^4 + \xi_{3}^4)|^2
\end{aligned}\label{expgtw}
\end{equation}
associated to the sixteen fixed loci of the $g$-twisted sector. The structure of the twisted vacua at each fixed point can be straightforwardly extracted from \eqref{expgtw} by keeping only the lightest characters $\xi_{\pm 1}$ and applying the map $\xi_1 \to \sigma_a$, $\xi_{-1} \to \sigma_{\dot a}$. For instance, the vacuum associated to the first monomial is
\begin{equation}
\prod_{i=1}^4 \sigma_a^i \tilde \sigma_a^i |0, \tilde 0\rangle\,.
\end{equation}

The amplitude ${\mathcal Z} \onech{g}{h}$ admitting the most general discrete torsion $\epsilon_r$ according to \cite{Gaberdiel:2004vx} is
\begin{equation}
\begin{aligned}
{\mathcal Z} \onech{g}{h} &= \left[\epsilon_{1} |(\xi_1^1 + \xi_{-3}^1)(\xi_1^2 + \xi_{-3}^2)(\xi_1^3 - \xi_{-3}^3)(\xi_1^4 - \xi_{-3}^4)|^2 \right.
\\
&\ \, + \epsilon_{2} |(\xi_1^1 + \xi_{-3}^1)(\xi_1^2 + \xi_{-3}^2)(\xi_1^3 - \xi_{-3}^3)(\xi_{-1}^4 - \xi_{3}^4)|^2
\\
&\ \, + \epsilon_{3} |(\xi_1^1 + \xi_{-3}^1)(\xi_1^2 + \xi_{-3}^2)(\xi_{-1}^3 - \xi_{3}^3)(\xi_1^4 - \xi_{-3}^4)|^2 
\\
&\ \, + \epsilon_{4} |(\xi_1^1 + \xi_{-3}^1)(\xi_{-1}^2 + \xi_{3}^2)(\xi_1^3 - \xi_{-3}^3)(\xi_1^4 - \xi_{-3}^4)|^2
\\
&\ \, + \epsilon_{5} |(\xi_{-1}^1 + \xi_{3}^1)(\xi_1^2 + \xi_{-3}^2)(\xi_1^3 - \xi_{-3}^3)(\xi_1^4 - \xi_{-3}^4)|^2 
\\
&\ \, + \epsilon_{6} |(\xi_1^1 + \xi_{-3}^1)(\xi_1^2 + \xi_{-3}^2)(\xi_{-1}^3 - \xi_{3}^3)(\xi_{-1}^4 - \xi_{3}^4)|^2
\\
&\ \, + \epsilon_{7} |(\xi_1^1 + \xi_{-3}^1)(\xi_{-1}^2 + \xi_{3}^2)(\xi_1^3 - \xi_{-3}^3)(\xi_{-1}^4 - \xi_{3}^4)|^2 
\\
&\ \, + \epsilon_{8} |(\xi_{-1}^1 + \xi_{3}^1)(\xi_1^2 + \xi_{-3}^2)(\xi_1^3 - \xi_{-3}^3)(\xi_{-1}^4 - \xi_{3}^4)|^2
\\
&\ \, + \epsilon_{9} |(\xi_1^1 + \xi_{-3}^1)(\xi_{-1}^2 + \xi_{3}^2)(\xi_{-1}^3 - \xi_{3}^3)(\xi_1^4 - \xi_{-3}^4)|^2 
\\
&\ \, + \epsilon_{10} |(\xi_{-1}^1 + \xi_{3}^1)(\xi_1^2 + \xi_{-3}^2)(\xi_{-1}^3 - \xi_{3}^3)(\xi_1^4 - \xi_{-3}^4)|^2
\\
&\ \, + \epsilon_{11} |(\xi_{-1}^1 + \xi_{3}^1)(\xi_{-1}^2 + \xi_{3}^2)(\xi_1^3 - \xi_{-3}^3)(\xi_1^4 - \xi_{-3}^4)|^2 
\\
&\ \, + \epsilon_{12} |(\xi_{-1}^1 + \xi_{3}^1)(\xi_{-1}^2 + \xi_{3}^2)(\xi_{-1}^3 - \xi_{3}^3)(\xi_1^4 - \xi_{-3}^4)|^2
\\
&\ \, + \epsilon_{13} |(\xi_{-1}^1 + \xi_{3}^1)(\xi_{-1}^2 + \xi_{3}^2)(\xi_{1}^3 - \xi_{-3}^3)(\xi_{-1}^4 - \xi_{3}^4)|^2
\\
&\ \, + \epsilon_{14} |(\xi_{-1}^1 + \xi_{3}^1)(\xi_{1}^2 + \xi_{-3}^2)(\xi_{-1}^3 - \xi_{3}^3)(\xi_{-1}^4 - \xi_{3}^4)|^2
\\
&\ \, + \epsilon_{15} |(\xi_{1}^1 + \xi_{-3}^1)(\xi_{-1}^2 + \xi_{3}^2)(\xi_{-1}^3 - \xi_{3}^3)(\xi_{-1}^4 - \xi_{3}^4)|^2 
\\
&\ \, +\left. \epsilon_{16} |(\xi_{-1}^1 + \xi_{3}^1)(\xi_{-1}^2 + \xi_{3}^2)(\xi_{-1}^3 - \xi_{3}^3)(\xi_{-1}^4 - \xi_{3}^4)|^2\right] 
\\
&\qquad \times \prod_{i \in \{5,6\}} \left[ |\xi^i_0 - \xi^i_4|^2 + |\xi^i_2 - \xi^i_{-2}|^2 \right]
\end{aligned}\label{Agrndt}
\end{equation}
The amplitude ${\mathcal Z}\onech{h}{g}$ is obtained via an $S$ modular transformation and, clearly, depends on the values of the $\epsilon_r$. A consistent set of values must be such that ${\mathcal Z}\onech{h}{g}$ defines a proper projection of ${\mathcal Z}\onech{h}{1}$, so to preserve the operator interpretation of the partition function. Clearly, $\epsilon_r=1$ or $\epsilon_r=-1$ for all $r$'s gives a consistent CFT, while other choices are forbidden. For instance, taking in \eqref{Agrndt} $\epsilon_1 = -1$ and all other equal to one, one obtains
\begin{equation}
\begin{aligned}
{\mathcal Z}\onech{h}{g} &=\left[ \frac{1}{8} (7 \, \xi_0^1 \xi_0^2 \xi_1^3 \xi_1^4  - i \, \xi_2^1\xi_0^2 \xi_1^3 \xi_1^4 )(\bar\xi_0^1 \bar\xi_0^2 \bar\xi_1^3 \bar\xi_1^4 ) + \text{65534\ extra\ terms} \right]\\
& \qquad \times \prod_{i \in \{5,6\}} \left[ |\xi^i_1 + \xi^i_{-3}|^2 + |\xi^i_{-1} + \xi^i_{3}|^2 \right]\,,
\end{aligned}
\end{equation}
which is clearly incompatible with ${\mathcal Z}\onech{h}{1}$. The same is true for other choices of (mixed) signs.

\section{The genus-two partition function for the \texorpdfstring{$S^1/\mathbb{Z}_2$}{S1/Z2} orbifold} \label{twoloopc1equiv}

We review here the main steps needed for the calculation of the genus-two partition function for the $T^6/\mathbb{Z}_2 \times \mathbb{Z}_2$ orbifold. Actually, we shall concentrate on a single real compact boson twisted by $\mathbb{Z}_2$, since it is the building block to construct the partition function for the full orbifold. 
As anticipated in Section \ref{sec:genus2}, an order two twist imposes periodic or anti-periodic conditions along the one-cycles of $\Sigma_2$. These are encoded in the twist vectors $\zeta_i$ of eq. \eqref{eq:twists}. The contribution of the trivial  twist $\zeta_0$ is quite simple since we are dealing with bosons periodic along all cycles, and is encoded in the Narain lattice $\Lambda_{1,1} (\Omega)$ of eq. \eqref{eq:g2narain}. More interesting is the contribution of the non-trivial twists with $i\neq 0$.  The standard procedure is to start from a suitable genus-three surface with periodic conditions for the compact boson. The Riemann surface $\Sigma_2$ with a twist vector $\zeta_i$ are then obtained via a anti-holomorphic involution $\iota_i$, acting as \cite{Dijkgraaf:1987vp}
\begin{equation}
   X(\iota _i (z), \overline{ \iota_i (z)}) =-X(z,\bar z)  + 2 \pi m R \, . 
\end{equation}
The solution of the Laplace equation with these boundary conditions is
\begin{equation}
    X(z,\bar z)=-\frac{i \pi R}{\sqrt{2\alpha'}}  \, (n-\bar \pi_i m) \int_{\iota_i (z)}^z\nu_i + \text{c.c} + X'(z,\bar  z) \, ,
\end{equation}
which has non-trivial zero modes along the Prym variety with Teichm\"uller parameter $\pi_i$ and quantum fluctuations $X'(z,\bar z)$. The zero-modes build the Narain lattice $\Lambda_{1,1} (R;\pi_i )$. 
The contribution of the quantum fluctuations to the genus-two partition functions is more involved, but fortunately, following \cite{Dijkgraaf:1987vp}, we can resort to a non-trivial property of $c=1$ systems, \emph{i.e.} the intersection of the moduli spaces of toroidal and orbifold compactifications \cite{Ginsparg:1987eb}. At this intersection point, the associated partition functions are then identical
\begin{equation}
    \mathcal{Z}_g^{S^1}(\sqrt{\alpha'}/2)= \mathcal{Z}_g^{S^1/\mathbb{Z}_2}(\sqrt{\alpha'}) \, ,
\end{equation}
which implies
\begin{equation}
 \Lambda_{1,1}(\sqrt{\alpha'}/2;\Omega)= \frac{1}{4} \left( \Lambda_{1,1}\left( \sqrt{\alpha'} ;\Omega\right) +\sum_{i=1}^{15}  |\mathcal{Z}_i|^2 \Lambda_{1,1}\left( \sqrt{\alpha'};\pi_i\right) \right) \, . \label{eq:c1equiv}
\end{equation}
The Narain lattices at those rational points can be conveniently written in terms of theta functions \cite{Dijkgraaf:1987vp}
\begin{equation}
\Lambda_{1,1} (\sqrt{\alpha'};\Omega)= \sum_{\rho\in \{ 0, \frac12\}^2} \left| \frac{\theta\onech{\rho}{0} (0|2\Omega)}{{\mathcal Z}_0 (\Omega )} \right|^2\, , 
\qquad
\Lambda_{1,1} \left(\sqrt{\alpha'};\pi_i\right)= \sum_{\rho=0,\frac{1}{2}} \left| \frac{\theta\onech{\rho }{ 0} (0|2\pi_i )}{\eta (\pi_i ) }\right|^2 \,,
\end{equation}
and
\begin{equation}
\Lambda_{1,1} (\sqrt{\alpha'}/2;\Omega) =\frac{1}{4}\, \sum_{\alpha , \beta, \gamma \in \{ 0, \frac12\}^2  } \left| \frac{\theta \onech{ \alpha + \beta/2}{\gamma} (0|2\Omega )}{{\mathcal Z}_0 (\Omega )}  \right|^2 \,.
\end{equation}
Inserting the latter expressions into eq. \eqref{eq:c1equiv}, one finds
\begin{equation}\label{eq:c1equivexplicit}
\begin{aligned}
     &\frac{1}{4}\, \sum_{\alpha , \beta, \gamma \in \{ 0, \frac12\}^2  } \bigg| \frac{\theta \onech{ \alpha + \beta/2}{\gamma} (0|2\Omega )}{{\mathcal Z}_0 (\Omega )}  \bigg|^2=
     \\
     &\qquad  \frac{1}{4} \Bigg ( \sum_{\rho\in \{ 0, \frac12\}^2} \bigg | \frac{\theta\onech{\rho}{0} (0|2\Omega)}{{\mathcal Z}_0 (\Omega )} \bigg|^2 +\sum_{i=1}^{15}  |\mathcal{Z}_i|^2  \sum_{\rho=0,\frac{1}{2}} \bigg| \frac{\theta\onech{\rho }{ 0} (0|2\pi_i )}{\eta (\pi_i ) }\bigg|^2 \Bigg) \, .
     \end{aligned}
\end{equation}
Notice that on the RHS the sum over all the possible {\em twists} can be written as
\begin{equation}
    \sum_{i=1}^{15} \qquad \to \qquad  \sum_{\substack{\lambda, \mu, \nu , \sigma=0,\frac12 \\ \neq (0,0,0,0)} } \, ,
\end{equation}
while a similar redefinition for the LHS 
\begin{equation}
    \sum_{\alpha , \beta, \gamma \in \{ 0, \frac12\}^2  } \left| \frac{\theta \onech{ \alpha + \beta/2}{\gamma} (0|2\Omega )}{{\mathcal Z}_0 (\Omega )}  \right|^2 = \sum_{\lambda, \mu , \nu , \rho, \sigma, \kappa=0,\frac12} \left| \frac{\theta \twoch{ \rho +\lambda/2}{\kappa+\mu/2}{\nu}{ \sigma} (0|2\Omega )}{{\mathcal Z}_0 (\Omega )}  \right|^2 \, ,
\end{equation}
includes odd spin structures, for which the theta function vanishes identically. We can therefore restrict the sum only to the even spin structures.
As a result,  eq. \eqref{eq:c1equivexplicit} becomes
\begin{equation}
\begin{aligned}
     &\frac{1}{4}\, \sum_{\lambda, \mu , \nu ,\sigma, \rho, \kappa=0,\frac12} \bigg|  \frac{\theta \twoch{ \rho +\lambda/2}{\kappa+\mu/2}{\nu}{ \sigma} (0|2\Omega )}{{\mathcal Z}_0 (\Omega )}  \bigg|^2
     \\
     &= \frac{1}{4} \Bigg ( \sum_{\rho,\kappa= 0, \frac12}  \bigg | \frac{\theta\twoch{\rho}{\kappa}{0}{0} (0|2\Omega)}{{\mathcal Z}_0 (\Omega )} \bigg|^2 +\sum_{\substack{\lambda, \mu, \nu , \sigma=0,\frac12 \\ \neq (0,0,0,0)} } |\mathcal{Z}_{(\lambda, \mu ,\nu , \sigma)}|^2  \sum_{\alpha=0,\frac{1}{2}} \bigg| \frac{\theta\onech{\rho }{ 0} (0|2\pi_{(\lambda, \mu ,\nu , \sigma)} )}{\eta (\pi_{(\lambda, \mu ,\nu , \sigma)} ) }\bigg|^2 \Bigg) \, ,
     \end{aligned}
\end{equation}
with suitable restrictions on the values of $\lambda, \mu , \nu ,\sigma, \rho, \kappa$ on the LHS. Actually, $\lambda$ and $\mu$ can take any value, and so can be assumed for $\nu$ and $\sigma$. Therefore, only $\rho$ and $\kappa$ are constrained. For instance, the contribution with $\nu=\sigma=\mu=\lambda=0$ leaves  $\rho$ and $\kappa$  unconstrained, and thus matches the first term on the RHS.
Next, we can take $\nu=\mu=\lambda=0$ and $\sigma=\frac12$, which fixes the twist $\zeta_1$. The LHS yields 
\begin{equation}
    \frac{1}{4}\, \sum_{ \rho, \kappa=0,\frac12} \bigg|  \frac{\theta \twoch{ \rho }{\kappa}{0}{ \nicefrac{1}{2}} (0|2\Omega )}{{\mathcal Z}_0 (\Omega )}  \bigg|^2=\frac{1}{4}\, \sum_{ \rho=0,\frac12} \bigg|  \frac{\theta \twoch{ \rho }{0}{0}{ \nicefrac{1}{2}} (0|2\Omega )}{{\mathcal Z}_0 (\Omega )}  \bigg|^2 \, ,
\end{equation}
since the choice $\kappa=\frac12$, for any $\rho$, gives an odd spin structure, see eq. \eqref{eq:oddspinstructures}.
One thus obtains
\begin{equation}
    \frac{1}{4}\, \sum_{ \rho=0,\frac12} \bigg|  \frac{\theta \twoch{ \rho }{0}{0}{ \nicefrac{1}{2}} (0|2\Omega )}{{\mathcal Z}_0 (\Omega )}  \bigg|^2=\frac14 |\mathcal{Z}_1|^2\sum_{\rho=0,\frac{1}{2}} \bigg| \frac{\theta\onech{\rho }{ 0} (0|2\pi_1 )}{\eta (\pi_1 ) }\bigg|^2
\end{equation}
which implies
\begin{equation}
{\mathcal Z}_1 = \frac{\theta \twoch{ \rho }{0}{0}{ \nicefrac{1}{2}} (0|2\Omega )}{{\mathcal Z}_0 (\Omega )}\, \frac{ \eta (\pi_1 ) }{\theta\onech{\rho }{ 0} (0|2\pi_1 )} \,  .
\end{equation}
Using the Schottky relations \eqref{eq:Schottky}, 
\begin{equation}
{\mathcal Z}_1^2 =   \frac{\theta\twoch{\alpha}{0}{\beta}{0} (0|\Omega)\, \theta\twoch{\alpha}{0}{\beta}{\nicefrac{1}{2}} (0|\Omega ) }{{\mathcal Z}_0 (\Omega )^2}\, \frac{\eta (\pi_1 )^2}
{\theta^2 \onech{\alpha}{\beta} (0|\pi_1 )} \, ,
\label{eq:Z1Schottky}
\end{equation}
valid for $\alpha$, $\beta$ arbitrary half-integer characteristics, with $(\alpha , \beta ) \neq (\nicefrac{1}{2},\nicefrac{1}{2})$. This gives three different, but equivalent, expressions for ${\mathcal Z}_1 ^2$.

The ${\mathcal Z}_i^2$, with $i \neq 1$, can then be obtained from  ${\mathcal Z}_1^2$ via the action of $\text{Sp} (4;\mathbb{Z})$ transformations. To this end, we notice that the period of the Prym variety has a non-trivial stabiliser inside the modular group. For $\pi_1$, the stabiliser $\text{Stab} (\pi_1)$ is generated by $S_1$, $T_1$ and $T_2$, since they act trivially on the twist $\zeta_1$ which, in turn, defines $\pi_1$.  Such transformations reproduce the arbitrariness of \eqref{eq:Z1Schottky}, which is a different way to justify the Schottky relations. An element $\gamma_i \in \text{Sp} (4;\mathbb{Z})/\text{Stab} (\pi_1 )$ acts linearly on $\zeta_1$ and maps it to the different twist $\zeta_i = \gamma_i \cdot \zeta_1$, as \cite{Aoki:2003sy}
\begin{equation}
\begin{pmatrix} \zeta^a_i \\ \zeta^b_i \end{pmatrix} = \begin{pmatrix} D_i & - C_i \\ -B_i & A_i \end{pmatrix} \, \begin{pmatrix} \zeta^a_1 \\ \zeta^b_1 \end{pmatrix}\,, \qquad \text{for}\qquad
\gamma_i = \begin{pmatrix} A_i & B_i \\ C_i & D_i \end{pmatrix} \,.
\end{equation}
As a result, $\pi_1 \to \pi_i$ via the M\"obius transformation associated to $\gamma_i$ which, however, leaves invariant the characteristics of the associated theta functions, $\theta \onech{\alpha}{\beta} (0|\pi_1 ) \to \theta \onech{\alpha}{\beta} (0|\pi_i )$. For instance, the Dehn twist $\Sigma '$ and the combination $\Sigma' S_1 T_2 \Sigma'$ map $\zeta_1$ to $\zeta_2$ and $\zeta_{12}$, respectively, and thus yield the expressions
\begin{equation}
{\mathcal Z}_2^2 = \frac{\theta\twoch{\alpha+\beta+\nicefrac{1}{2}}{\beta+\nicefrac{1}{2}}{\beta}{0} (0|\Omega)\, \theta\twoch{\alpha+\beta}{\beta}{\beta}{\nicefrac{1}{2}} (0|\Omega)}{{\mathcal Z}_0  (\Omega)^2}\, 
\frac{\eta (\pi_2)^2}{\theta\onech{\alpha}{\beta} (0|\pi_2)^2}\,,
\end{equation}
and 
\begin{equation}
{\mathcal Z}_{12}^2 =\frac{\theta\twoch{\alpha+\beta}{\alpha+\beta+\nicefrac{1}{2}}{\alpha+\beta+\nicefrac{1}{2}}{\beta}  (0|\Omega)\, 
\theta\twoch{\alpha+\beta+\nicefrac{1}{2}}{\alpha+\beta+\nicefrac{1}{2}}{\alpha+\beta}{\beta}  (0|\Omega)}{{\mathcal Z}_0 (\Omega) ^2}\, 
\frac{\eta (\pi_{12})^2}{\theta\onech{\alpha}{\beta} (0|\pi_{12})^2}\,,
\end{equation}
up to an overall phase which cancels out in the partition function. Similarly, one can obtain the quantum contributions associated to the other twists, which altogether build the full partition function.

\begin{table}
\centering
\begin{tabular}{c|ccccccccccccccc}
    $i$ & 1 & 2 & 3 & 4 & 5 & 6 & 7 & 8 & 9 & 10 & 11 & 12 & 13 & 14 & 15
    \\
    \hline
    \hline
    $\delta_i (3)$ & 1 & 4 & 9 & 9 & 4 & 2 & 1 & 1 & 5 & 6 & 5 & 6 & 2 & 1 & 1
    \\
    \hline
    $\delta_i (4) $ & 3 & 7& 5 & 5 & 7 & 7 & 7 & 7 & 4 & 2 & 4 & 2 & 9 & 7 & 7
    \\
    \hline
    $\delta_i (2)$ & 8 & 10 & 6 & 6 & 2 & 10 & 8 & 3 & 6 & 1 & 1 & 3 & 10 & 3 & 8
\end{tabular}
\caption{The association between the twists $\zeta_i$ and the even spin structures $\delta_i (\alpha)$ of the genus-two theta functions. The argument $\alpha$ refers to the genus-one Jacobi theta functions $\vartheta_\alpha$ which enter the two-loop vacuum amplitude. The ambiguity in the choice of the even spin structures reflects the Schottky relations which yield different, but equivalent, expressions for the ${\mathcal Z}_i$.}
\label{tab:spinstructuresSchottky}
\end{table}
 
The Schottky relations provide different, but equivalent, expressions for the full genus-two vacuum energy of the $T^6 /\mathbb{Z}_2^2$ orbifold. They generalise table \ref{tab:spinstructures} to table \ref{tab:spinstructuresSchottky}.

The last step in the construction and analysis of the genus-two amplitude is the discussion of the factorisation limit. To this end, we parametrise the period matrix as in eq. \eqref{eq:Omegaparametrisation} and study the $t\to 0$ limit.  The genus-two theta functions are easy to handle since they obey eq. \eqref{eq:thetafactorisation}. Similarly, ${\mathcal Z}_0 (\Omega ) \to \eta (\tau_1) \, \eta (\tau_2 )$. 
The behaviour of the periods of the Prym variety, and that of the theta functions and Narain lattice built upon it, is less straightforward and, to this end, the Schottky relations come to our rescue. Let us focus, once more, on the period $\pi_1$ associated to the twist $\zeta_1$. From eq. \eqref{eq:Z1Schottky} we can write
\begin{equation}
    \left( \frac{\theta \twoch{0}{0}{0}{0}(0|\Omega) \ \theta \twoch{0}{ 0}{  0}{ \nicefrac{1}{2}}(0|\Omega)}{  \theta  \onech{0 }{0}(0|\pi_1)^2} \right)^2=\left( \frac{\theta \twoch{0}{  0}{ \nicefrac{1}{2}}{  0}(0|\Omega) \ \theta \twoch{0}{  0}{  \nicefrac{1}{2}}{  \nicefrac{1}{2}}(0|\Omega)}{    \theta \onech{0}{ \nicefrac{1}{2}}(0|\pi_1)^2} \right)^2\, , 
\end{equation}
and, using the factorisation properties of the genus-two theta functions \eqref{eq:thetafactorisation}, one gets
\begin{equation}
\left( \frac{\theta\onech{0}{  0}(0|\tau_1)}{\theta\onech{0}{ \nicefrac{1}{2}}(0|\tau_1)} \right)^4  \sim \left( \frac{\theta\onech{0 }{ 0}(0|\pi_1)}{\theta\onech{0 }{ \nicefrac{1}{2}}(0|\pi_1)} \right)^4 \, .
\end{equation}
As a result, $\pi_1 = \tau_1 + O(t)$.

For other periods we can proceed in a similar way, although, at times, the solution can be more complicated. For instance, for $\zeta_2$ the Schottky relations 
\begin{equation}
    \left( \frac{\theta \twoch{0}{  0 }{ 0 }{ \nicefrac{1}{2}}(0|\Omega) \ \theta \twoch{\nicefrac{1}{2}}{  \nicefrac{1}{2}}{ 0}{  0}(0|\Omega)}{   \theta \onech{0 }{ 0}(0|\pi_2)^2} \right)^2=\left(  \frac{\theta \twoch{\nicefrac{1}{2}}{  \nicefrac{1}{2}}{\nicefrac{1}{2}}{  \nicefrac{1}{2}}(0|\Omega) \ \theta \twoch{0}{  0}{ \nicefrac{1}{2}}{  0}(0|\Omega)}{ \theta \onech{0}{ \nicefrac{1}{2}}(0|\pi_2)^2} \right)^2 \,,
\end{equation}
imply, in the factorisation limit,
\begin{equation}
\left( \pi^2 t \  \theta\onech{0 }{ \nicefrac{1}{2}}(0|\tau_1)^2 \theta\onech{0 }{ 0}(0|\tau_2)^2\right)^2  \sim \left( \frac{\theta\onech{0 }{ \nicefrac{1}{2}}(0|\pi_2)}{\theta\onech{0 }{ 0}(0|\pi_2)} \right)^4 \,.
\end{equation}
Therefore, the Prym period $\pi_2$ must vanish as $t\to 0$. To better parametrise the solution we perform an $S$ modular transformation to get
 \begin{equation}
    \pi_2 \sim -\frac{\pi i}{\log \left[ (\pi/2)^4 \, t^2 \,  \theta\onech{0}{ \nicefrac{1}{2}}(0|\tau_1)^4\, \theta\onech{0 }{ 0}(0|\tau_2)^4\right]} \, .
\end{equation}
Similarly, one can derive the behaviour of the Prym periods $\pi_i$, associated to the other twists $\zeta_1$, in the factorisation limit.

Of course, one has to consider also the $t\to 0$ limit of the Narain lattice build upon the Prym variety. For the twist $\zeta_1$ things are quite simple: $\pi_1 \to \tau_1$ and, therefore, $\Lambda_{1,1} (\pi_1 ) \to \Lambda_{1,1} (\tau_1 )$. Less straightforward is the situation for $\Lambda_{1,1} (\pi_2 ) $. In this case, to derive the behaviour of the Narain lattice as $\pi_2\to 0$ it is convenient to Poisson resum over both momenta and windings, which effectively sends $\pi_2 \to -1/\pi_2$. As a result, only the term with vanishing momenta and windings survives the limit, and 
\begin{equation}
\frac{\eta (\pi_2 ) }{ \theta \onech{0}{0} (0|\pi_2 )}\,\Lambda_{1,1} (\pi_2 )   \to 1 \qquad \text{as} \qquad \pi_2 \to 0\,.
\end{equation}

\newpage


 \bibliographystyle{JHEP}
 \bibliography{String_biblio.bib}

\end{document}